\documentclass[%
 reprint,
 amsmath,amssymb,
 aps]{revtex4-2}

\usepackage{graphicx}
\usepackage{dcolumn}
\usepackage{float}
\usepackage{xcolor}
\usepackage{bm}
\usepackage{}
\usepackage{gensymb}
\begin{document}
\preprint{APS/123-QED}
\title{Enhanced Frequency noise suppression for LISA by combining cavity and arm locking control systems}
\author{Jobin Thomas Valliyakalayil$^{1}$}
\email{JobinThomasValliyakalayil@anu.edu.au}
\author{Andrew J. H. Sutton$^1$}
\author{Robert E. Spero$^2$}
\author{Daniel A. Shaddock$^1$}
\author{Kirk McKenzie$^{1}$}

\affiliation{ 1. Centre for Gravitational Astrophysics, Australian National University\\
2. Jet Propulsion Laboratory, California Institute of Technology}
\date{\today}

\begin{abstract}
This paper presents a novel method for laser frequency stabilisation in the Laser Interferometer Space Antenna (LISA) mission by locking a laser to two stable length references – the arms of the interferometer and an on-board optical cavity. The two references are digitally fused using carefully designed control systems, attempting minimal or no changes to the baseline LISA mission hardware. The interferometer arm(s) provides the most stable reference available in the LISA science band (0.1 mHz - 1 Hz), while the cavity sensor's wide-band and linear readout enables additional control system gain below and above the LISA band. The main technical issue with this dual sensor approach is the undesirable slow laser frequency pulling which couples into the control system with the imperfect knowledge of the Doppler shift of the light due to relative spacecraft motion along the LISA arm. This paper outlines requirements on the Doppler shift knowledge to maintain the cavity well within the resonance when activating the fused control system. Two Doppler shift estimation methods are presented that use the already on-board measurements, the inter-spacecraft interferometer link (the main science measurement), and the absolute inter-spacecraft laser ranging system. Both methods reach the required precision after a few thousand seconds of measurement integration. The paper demonstrates an approach to initialise and engage the proposed laser stabilization system, starting from free-running laser and ending with the dual sensor frequency control system. The results show that the technique lowers the residual laser frequency noise in the LISA science band by over 3 orders of magnitude: from 30 Hz/$\sqrt{\textrm{Hz}}$ to as low as 7 mHz/$\sqrt{\textrm{Hz}}$, potentially allowing the requirements on Time-Delay-Interferometry (TDI) to be relaxed - possibly to the point where first-generation TDI may be sufficient. 

\end{abstract}

\maketitle

\section{\label{Introduction}Introduction}

The Laser Interferometer Space Antenna  (LISA) mission is a space-based gravitational wave detector, proposed and planned jointly by ESA, NASA and a European consortium~\cite{LISAL3,PPA}. The LISA mission consists of three spacecrafts in a triangular formation with 2.5~million kilometer arms. The spacecraft exchange laser beams, employing heterodyne interferometry to measure displacement between  spacecraft to detect gravitational waves in the frequency band between 0.1 mHz to 1 Hz, a frequency band inaccessible to ground-based detectors. The displacement sensitivity goal should be less than 10 pm/$\sqrt{\textrm{Hz}}$ for each arm-link to reach a strain sensitivity of 10$^{-21}$/$\sqrt{\textrm{Hz}}$~\cite{LISAL3}.

Stabilisation of laser frequency noise is critical for LISA to meet the design sensitivity, as it is indistinguishable from displacement along a single link~\cite{PPA}. LISA will lock the laser to a fixed-length ultra-stable optical cavity as reference using the Pound-Drever-Hall (PDH) technique~\cite{PDH}. A similar technique has been recently demonstrated using the laser instrument on the GRACE Follow-On mission~\cite{gracefo_cavity} demonstrating the required cavity performance. However stabilisation alone cannot realise the LISA sensitivity requirement, hence Time Delay Interferometry (TDI)~\cite{TDI,Post_TDI,TDI_exp} will be employed. TDI is a post-processing scheme that synthesizes laser frequency noise free measurement variables by forming two-beam, equal arm-length interferometer combinations using algebraic combinations of delayed link displacement measurements. Velocity correcting TDI combinations, also known as second-generation TDI~\cite{TDI-2}, and the Sagnac combination~\cite{TDI1.5} overcome the effect of spacecraft motion during the light travel time. In addition to TDI, laser stabilisation is required to a level of 282 Hz/$\sqrt{\textrm{Hz}}$ for second-generation TDI and to 1 Hz/$\sqrt{\textrm{Hz}}$ for first-generation TDI~\cite{mdual,FCST}. While second-generation TDI is currently being subject to detailed analysis to ensure all signal and noise propagation is understood~\cite{TDIn-2}, experimental  validation of  TDI~\cite{TDI_exp}  on the ground is extremely challenging. A supplementary method of laser stabilisation, such as Arm Locking, will provide risk reduction against unknowns and  margin to link performance.
   
LISA's inter-satellite link is designed (and required) to provide an intrinsic length stability below the $\sim10~\textrm{pm}/\sqrt{\textrm{Hz}}$ noise floor in order to detect gravitational waves~\cite{LISAL3}. Hence, by design, the LISA arm itself provides a highly stable length reference over frequencies of interest. Arm Locking was proposed by Sheard et al.~\cite{SingleArm} to use the arm itself for laser frequency noise stabilization and to provide further margin for TDI operation, which can be demonstrated by the GRACE-FO mission as proposed in~\cite{arm_grace}. Later  a dual arm sensor  was developed~\cite{dualarm,HansReiner} that utilised the two arm lengths with each spacecraft, and allowed more freedom in the design of the controller by moving the nulls to larger frequencies. This sensor was prone to \textit{Doppler pulling}~\cite{exp_mdual}, whereby on every light round trip time (16.7~s for single arm locking, significantly less for dual arm locking), the laser frequency experienced a shift, equal to the error in the knowledge of the Doppler shifts. Doppler pulling was reduced with the help of modified-dual arm locking~\cite{mdual,arm_thorpe}, that combined both common and dual arm locking sensors. However none of these arm locking systems were compatible with the LISA hardware baseline: cavity stabilization to a fixed-length resonator. To date, to maintain compatibility of arm-locking with an optical cavity the approaches investigated required a tunable cavity length~\cite{tunablespacer} or employ modulation-frequency that could be tuned~\cite{thorpe}.

This paper re-examines a combination of cavity locking (using the baseline fixed resonator length) and arm locking, and shows
\begin{itemize}
  \item Suppression of laser frequency noise below the requirement-level cavity noise (over 3 orders of magnitude of suppression),
   \item Successful acquisition and operation of cavity lock in the presence of Doppler pulling, so long as a prior Doppler estimate is available.
\end{itemize} 

We emphasise that this combination relaxes the TDI suppression requirement with minimal or no change to the optical or electronic hardware on board the spacecraft - relying mostly on an FPGA upload to existing digital hardware and/or a flight software update. Even if arm locking is not considered as baseline for laser frequency stabilisation, this implementation could allow for deployment after the launch of the spacecraft.   The difficulty faced for this hardware simplification is the requirement for high accuracy of knowledge of  Doppler shift. This paper studies this resultant laser frequency pulling and outlines the requirements for robust lock acquisition and stable operation thereafter.  
   
The paper is divided into 7 sections. Section~\ref{Model} discusses the system model used for  analysis and simulations, including the noise propagation through the system. Section~\ref{Doppler_data} discusses Doppler pulling with respect to the current model and~\ref{Timescales} explores the convergence times for several Doppler estimation techniques to measure the Doppler shifts. Section~\ref{Controller_design} presents the controller design for the combined arm and cavity references, while Section~\ref{Simulink_model} discusses a Simulink model used to verify the system in the time domain. Section~\ref{Discussion} looks into some insights for utilising the combination of sensors, while Section~\ref{Conclusions} gives the conclusions and possible future scope of this technique.

\section{\label{Model}Model}

\begin{figure*}[t!]
\includegraphics[width=18cm]{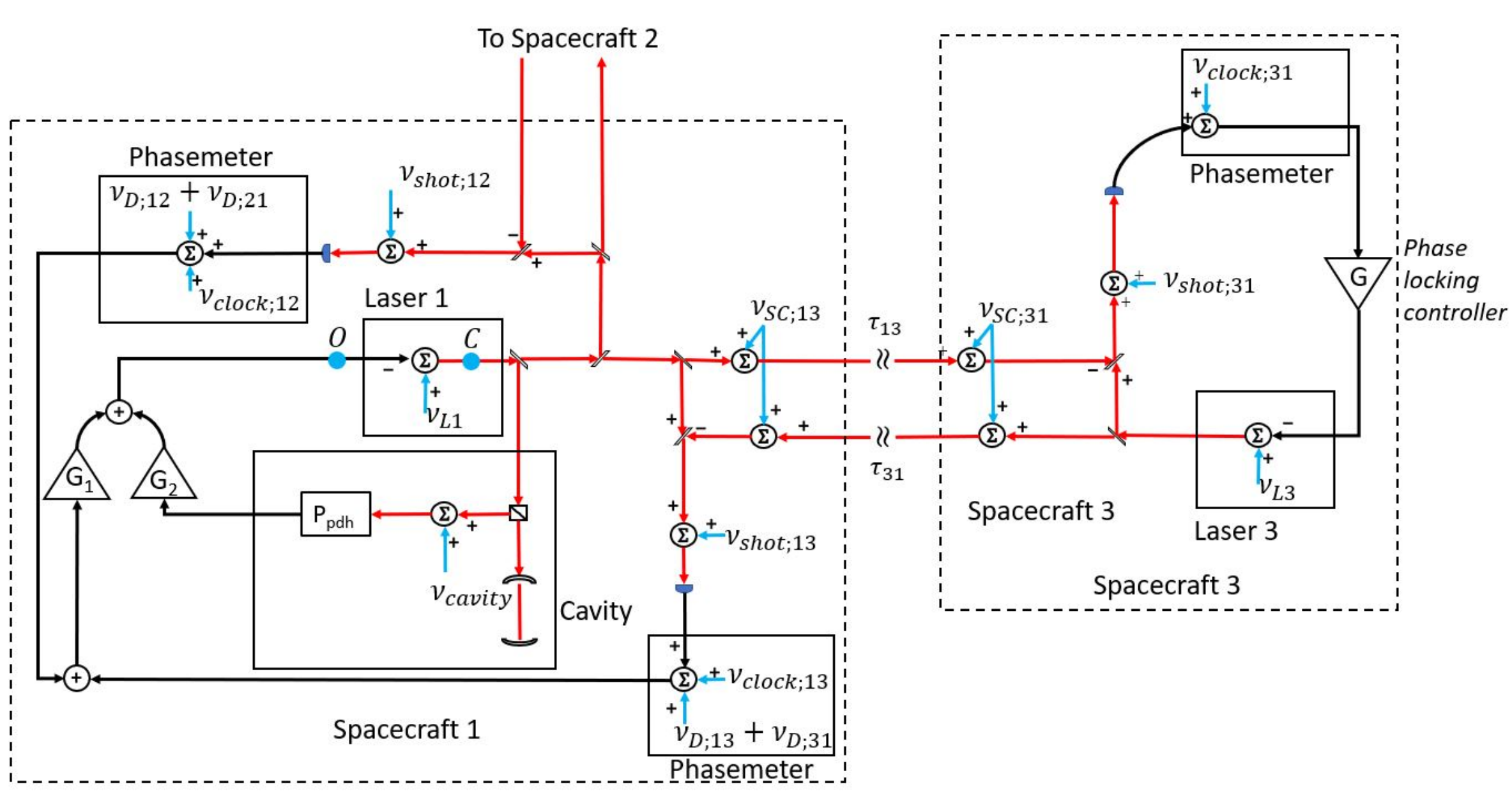}
\caption{Simplified model of the feedback system using Spacecraft 1 and 3 of LISA. Spacecraft 2 is not shown in this model, but can be considered symmetric to Spacecraft 3. $\tau_{13}$ and  $\tau_{31}$ are the travel times taken for the laser to traverse the separation between Spacecraft 3 and 1. The laser frequency noise, $\nu_{L_i}$, the clock noise, $\nu_{clock;ij}$, the shot noise, $\nu_{shot;ij}$, the spacecraft motion noise, $\nu_{SC;ij}$, the Doppler shifts, $\nu_{D;ij}$, the cavity noise, $\nu_{cavity}$ are the noise sources that are considered in this model. The indices i and j takes the values of 1, 2 or 3 referring to the different spacecrafts. The red trace outlines the path of the laser in free space, while the black trace of the path are in electrical/digital hardware.}
\label{Model_n}
\end{figure*}

Figure~\ref{Model_n} is the schematic diagram of the model that includes the main sensors, controllers, and the noise sources. All the equations in this paper are derived with respect to this model. The laser frequency noise, from the laser source in Spacecraft 1, serves as the main input and is fed into an arm sensor and the cavity(PDH) sensor. The output of these sensors, including the noise contributions, are given to individual controllers. The output from the two controllers are summed and fed back to the laser source and adjusts the laser frequency through PZT and thermal actuators, and thus create a feedback system. The notations and formalism used in this paper follow McKenzie et al.~\cite{mdual}. 

\subsection{\label{Sensors}Sensors}

{From previous arm locking analysis, the frequency noise sensors can be styled as a single arm sensor~\cite{SingleArm}, common arm sensor~\cite{dualarm}, difference arm sensor~\cite{dualarm}, dual arm sensor~\cite{dualarm} or modified-dual arm sensor~\cite{mdual}. In this paper, the common arm sensor is used, in which the average displacement measurements from two interferometer arms are used to measure laser frequency noise.For simplicity a single arm sensor could also be chosen, and we expect only minor modifications to the results. We approximate the high gain transponder systems of the other spacecrafts to act as a active retro-reflector, that transponds time-delayed laser phase, and hence simplify computations. The common arm locking transfer function from laser frequency to sensor (displacement) output is given by;}
\begin{equation}\label{ArmSensor}
    P_{+} (s)= {S_+}
    \begin{bmatrix}
        P_{12}(s)\\
        P_{13}(s)
    \end{bmatrix}   ,  \hspace{1cm} 
   {S_+} =  \begin{bmatrix}
      1 & 1
\end{bmatrix}.
\end{equation}
$S_+$ refers to the sensor matrix that utilises the individual arm responses~\cite{mdual}. The frequency response of the individual arms is given by

\begin{equation}
    P_{12}(s)=1-e^{-2s\tau_{12}} ,\hspace{0.5cm}
    P_{13}(s)=1-e^{-2s\tau_{13}},
\end{equation}
where $\tau_{12}$ and $\tau_{13}$ are the times for the laser-light to traverse arm 1 (from Spacecraft 1 to 2) and arm 2 (from Spacecraft 1 to 3) and relates to the arm lengths, $L_{1}$ and $L_{2}$ as $\tau_{12}=L_{1}/c$ and $\tau_{13}=L_{2}/c$ (c is the speed of light). For simplicity, the arm lengths are assumed to be constant and symmetric within the round trip time of the laser and thus have the approximations  $\tau_{21}$ = $\tau_{12}$ and $\tau_{31}$ = $\tau_{13}$. The parameter $s$ refers to the complex variable in Laplace domain. Equation~\ref{ArmSensor} can be also written analytically as
 
\begin{equation}
    P_+ (s)=2(1-\cos({\omega\Delta\tau})e^{-s\Bar{\tau}}),    
\end{equation}
where $ \space \bar{\tau} = \tau_{12}+\tau_{13}$ and  $\Delta\tau = \tau_{12}-\tau_{13}$. Here the value of $\bar{\tau} = 16.67$~s and the value of $\Delta\tau = 0.083$~s. This is with consideration the average length of the arms to be 2.5 million kms and the maximum difference in the lengths to be 1\% of that length~\cite{Orbits}. This approximation may not be valid during the entire LISA mission at which the phase margin may decrease by 1$\degree$ in the worst case scenario, when the arm lengths are equally matched, which has negligible effect on the results in this paper. Prior to the controllers, the sensors are scaled with a normalisation factor of $\frac{1}{2}$.

The cavity PDH locking sensor is modelled as a low-pass filter whose cut-off frequency, $f_c$, is at the cavity half-width half-max(HWHM) frequency for a high finesse cavity~\cite{EvanH}. With $f_c$ referring to the cavity pole, the PDH sensor can be shown as
\begin{equation}\label{PDHSensor}
    P_{\textrm{pdh}} (s)=\frac{D_0}{1+\frac{s}{2\pi f_c}}.
\end{equation}

$D_0$ is a gain that is dependent on the length of the cavity, the reflective and loss coefficients of the mirrors and the power modulation coefficients in the PDH scheme~\cite{EvanH}. We normalise the PDH sensor gain $D_0 = 2$ to be compatible with the arm sensor (In practice, this will require scaling the error signal prior to the controller). For LISA, the cavity on the spacecraft is expected to have a length of 7.77 cm with a free spectral range of approximately 2 GHz, similar to the cavity that is used in GRACE Follow-On mission~\cite{gracefo_cavity,grace_cavity}. Hence for a finesse of 10,000, the HWHM frequency can be computed to be approximately $f_c = 100$~kHz.
\begin{figure}[H]
\hspace*{-0.55cm}
   \includegraphics[width=7.15cm, height=10.2cm, angle=-90]{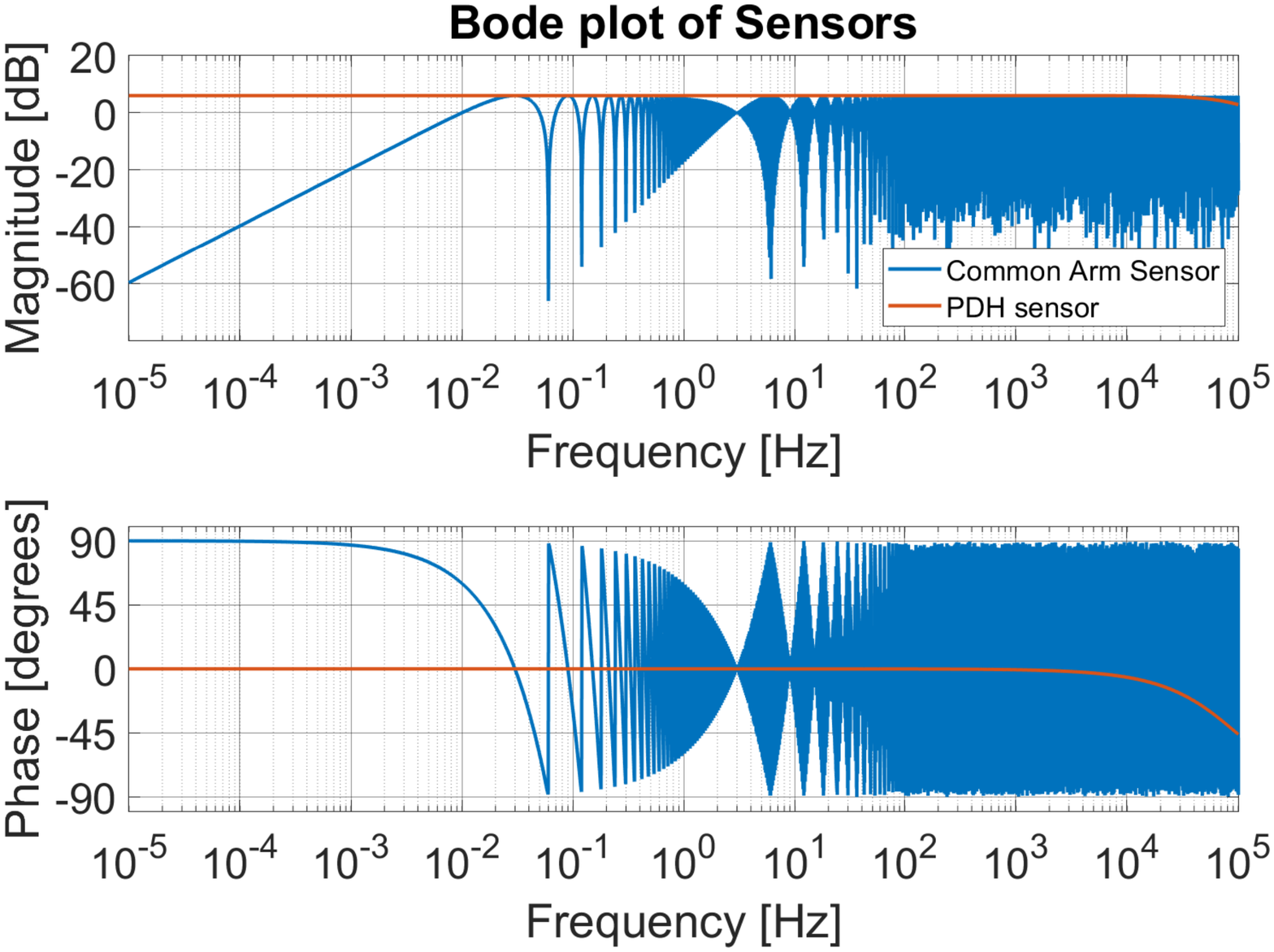}
  \caption{Bode plot of the arm sensor and the PDH sensor. The arm sensor is modelled as the common sensor considered with $\bar{\tau} = 16.67$~s  and $\Delta\tau = 0.08335$~s. The arm sensor has nulls at multiples of 60 mHz where the phase goes from -90$\degree$ to 90$\degree$. The difference in the arm lengths is reflected in the null depths. The PDH sensor has a flat response with the corner frequency at approximately 100 kHz, and a phase of $-6\degree$ at 10 kHz}
  \label{Sensor}
  \end{figure}
  \vspace{1cm}
The bode plot of both sensors are shown in Figure  \ref{Sensor}. The basic design of the controller presented in Section~\ref{Controller_design} relies on the  arm sensor being dominant (higher gain) in the LISA science band ($10^{-4}$~Hz to 1 Hz) since it offers the best frequency stability. Outside the LISA science band the controller design emphasizes the cavity sensor; at frequencies below $10^{-4}$~Hz to limit the laser frequency pulling due to imperfect Doppler shift knowledge, and above 1 Hz to increase the control system's phase margin at high frequencies and through the nulls of the common arm sensor. The unity gain frequency of the controller is selected at approximately 10 kHz, a decade lower than the cavity HWHM frequency, to utilise the flat response of the PDH sensor.

\subsection{\label{Noise_sources}TDI Noise Requirements}

The amount of laser frequency noise that TDI can suppress is limited by the time synchronization errors~\cite{TDI}. Dedicated inter-spacecraft ranging, using pseudo random noise modulation~\cite{PRN_ref,PRN_ref2}, is proposed to provide a spacecraft ranging error of 1~m and with an allowed residual noise floor of 2 pm/$\sqrt{\textrm{Hz}}$, the laser noise requirement before second-generation TDI is applied~\cite{FCST} is 
        \begin{equation}\label{NSF}
            \Tilde{\nu}_{\textrm{TDI-2}}(f)  = 282 \frac{\textrm{Hz}}{\sqrt{\textrm{Hz}}} . \sqrt{1+\left( \frac{2\hspace{0.05cm}\textrm{mHz}}{f} \right)^4}.
        \end{equation}
        
 First-generation TDI on the other hand does not correct for errors in the lengths due to the relative velocity between the spacecrafts. If the ranging is considered to be constrained by relative velocity of $\pm5$ ms$^{-1}$~\cite{LISAL3} between the spacecrafts, then in the worst case scenario, both the spacecrafts would experience a maximum relative velocity of 10 ms$^{-1}$, resulting in a change in range of 167~m in one round-trip of the laser (around 16.67~s). Thus, the laser noise suppression requirement prior to first-generation TDI  is
        \begin{equation}\label{NSF2}
            \Tilde{\nu}_{\textrm{TDI-1}}(f) = 1.7 \frac{\textrm{Hz}}{\sqrt{\textrm{Hz}}} . \sqrt{1+\left( \frac{2\hspace{0.05cm}\textrm{mHz}}{f} \right)^4}.
        \end{equation}
While second-generation TDI is a conceptual next step from first-generation, the noise couplings of different noise sources in second-generation TDI may be more complex due to the time-varying arm-lengths~\cite{TDIn-2,TDIn-3}. 

As a goal, this paper designs an arm-locking/cavity control system with ability to reduce the laser frequency noise to below the required level for application of first-generation TDI. This additional noise reduction results in additional margin for the the required TDI suppression and could be seen as risk reduction for second-generation TDI, and if needed may mean that first-generation TDI would be sufficient.
\subsection{
\label{Noise_prop}Noise propagation}
This subsection explains the propagation of the various noise sources as shown in Figure~\ref{Model_n}. The noise sources, notations, and transfer functions  follows the previous arm locking literature~\cite{SingleArm,dualarm,exp_mdual,mdual} and are described in Appendix A. In the following analysis, we use the notation of laser frequency noise (Hz/$\sqrt{\textrm{Hz}}$) as opposed to laser phase noise (cycles/$\sqrt{\textrm{Hz}}$) in other related papers and drop the Laplace `(s)' notation from the noise models and controllers in this subsection. For instance, $\nu_O(s)$ will be denoted as $\nu_O$.

\begin{figure}[H]
    \hspace*{-0.6cm}
    \includegraphics[width=7.25cm,height=10.25cm,angle=-90]{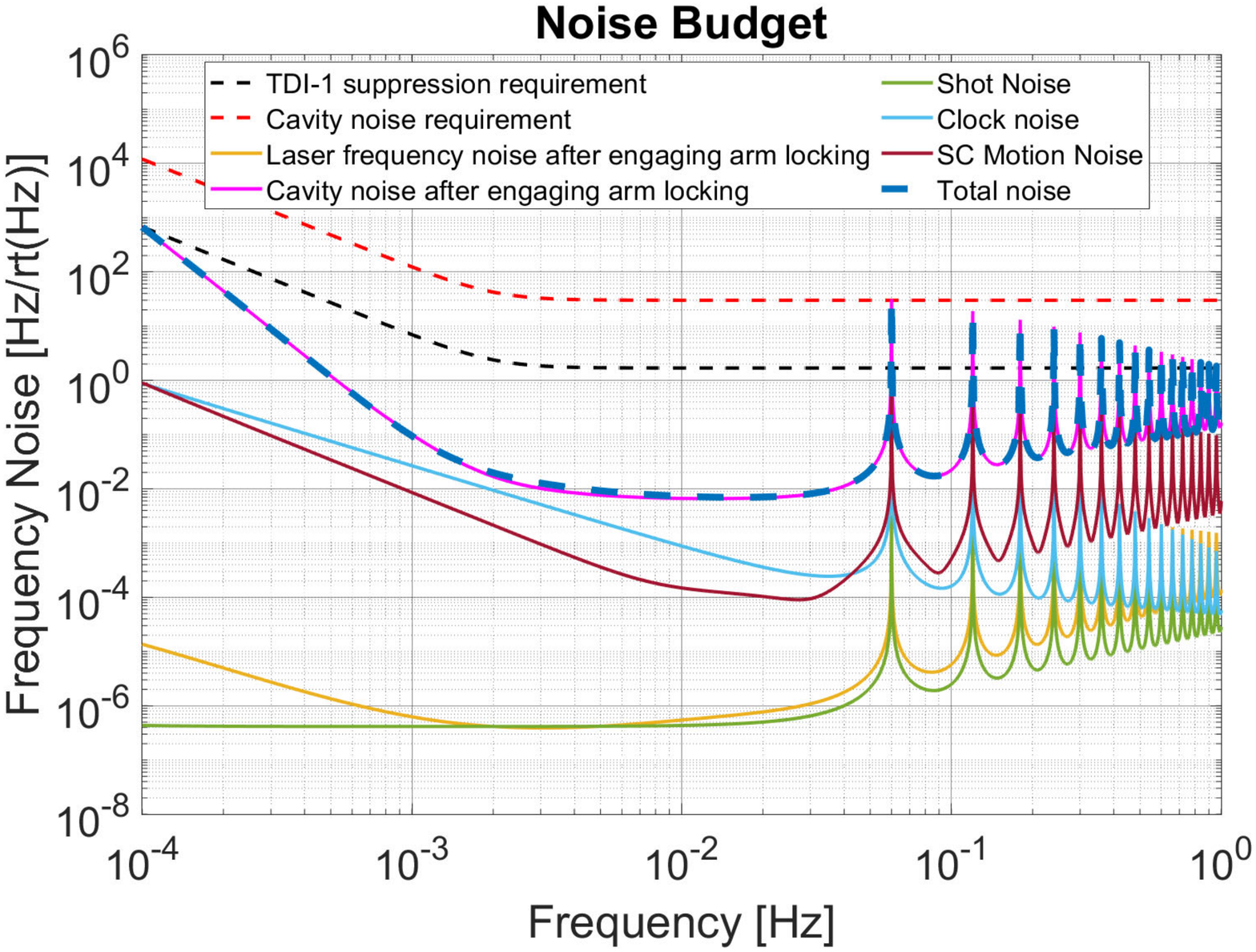}
    \caption{Noise budget using the noise models and the transfer functions along with the LISA requirements. The main contributing noise source is the requirement-level cavity noise that is suppressed from engaging the arm locking controller (pink trace) and within the first-generation TDI requirement (dashed black line) in the science band.}
    \label{fig:noise_budget}
\end{figure}

The laser frequency noise, $\nu_L$, is added at the laser source, and propagates through the interferometer and the cavity experiencing the transfer functions in Equations~\ref{ArmSensor} and~\ref{PDHSensor}. The requirement-level cavity noise, $\nu_{cavity}$, gets added into the system as the base stabilisation~\cite{PPA} provided by PDH locking and hence will propagate when the PDH error is taken. This requirement-level cavity noise will be dominated by effects such as readout noise and Brownian thermal noise. Shot noise, $\nu_{shot;ij}$, is added at photodetector in Spacecraft \textit{i}, when interfered with a laser from Spacecraft \textit{j}. There are
four independent shot noise contributions that are added at the primary spacecraft. Clock noise couples in each of the phasemeter reading between spacecraft 1 and spacecrafts 2 and 3, totalling to four terms of clock noise, $\nu_{clock;ij}$, with the beat-note frequency being the maximum value (25 MHz)~\cite{LISAL3}. The spacecraft motion noise, $\nu_{SC;ij}$, get coupled into the system whenever there is a change in the link between two spacecrafts and hence eight terms are contributed by this noise~\cite{mdual}. The total noise propagation through the entire open loop system at point O in Figure~\ref{Model_n} can be shown as:
    \begin{equation}\label{NoiseMatrix}
        \begin{split}
        \nu_O&= G_1 S_+ \begin{bmatrix}
            N_L+N_{SN}+N_{CN}+N_{SCN} 
        \end{bmatrix} 
        \\ &+ G_2 P_{\textrm{pdh}}\nu_{L} + G_2 P_{\textrm{pdh}}\nu_{cavity},
        \end{split}
    \end{equation}
where $N_L$ is the laser noise sensed at the primary spacecraft photodetectors, $N_{SN}$ is the shot noise, $N_{CN}$ is the clock noise,
and $N_{SCN}$ is the spacecraft motion noise. These are given by
    \begin{equation}\label{Noisematrix2}
    \begin{split}
       N_L &= \begin{bmatrix}
             P_{12}\nu_{L}\\
             P_{13}\nu_{L}
        \end{bmatrix} \\
        N_{SN} &=\begin{bmatrix}
             \nu_{shot;12} + \nu_{shot;21}e^{-s\tau_{12}}\\
             \nu_{shot;13} + \nu_{shot;31}e^{-s\tau_{13}}
        \end{bmatrix}\\
        N_{CN} &= \begin{bmatrix}
                \nu_{clock;12} + \nu_{clock;21}e^{-s\tau_{12}}\\
             \nu_{clock;13} + \nu_{clock;31}e^{-s\tau_{13}}
        \end{bmatrix}\\
        N_{SCN} &=\begin{bmatrix}
             -\nu_{SC;12}\left(1+e^{-2s\tau_{12}}\right)-2\nu_{SC;21}e^{-s\tau_{12}}\\
             -\nu_{SC;13}\left(1+e^{-2s\tau_{13}}\right)-2\nu_{SC;31}e^{-s\tau_{13}} 
        \end{bmatrix}.\\
        \end{split}
    \end{equation}
G$_1$ and G$_2$ are controllers used for each of the arm sensor and cavity sensor, respectively. Using equations~\ref{NoiseMatrix} and~\ref{Noisematrix2}, with transfer functions~\ref{T1}, ~\ref{T2}, ~\ref{T3}, ~\ref{T4}, ~\ref{T5}, the complete closed loop equation at point C can be written as
    \begin{equation}\label{Totalnoise}
    \begin{split}
       &\nu_{C}=  \nu_{L} - \nu_O \\
       =& \frac{\nu_{L}}{1+G_1 P_+ + G_2 P_{\textrm{pdh}}} - \frac{\nu_{cavity} G_2P_{\textrm{pdh}}}{1+G_1 P_+ + G_2 P_{\textrm{pdh}}} 
        \\
       -&\frac{G_1}{1+G_1 P_+ + G_2 P_{\textrm{pdh}}} S_+
       \begin{bmatrix}
             N_{SN}+N_{CN}+N_{SCN} 
       \end{bmatrix}. \\
        \end{split}
    \end{equation}
     The noise budget after engaging the hybrid control system is plotted in Figure~\ref{fig:noise_budget} using controllers for each sensor (covered in detail in Section~\ref{Controller_design}). The black dashed line plots the pre-TDI requirements based on the LISA sensitivity goal (Equation~\ref{NSF2}). It can be seen that the requirement-level cavity noise (shown as red dashed lines) is suppressed by over 3 orders at 10 mHz (shown as pink line) and is the dominant noise in the noise budget. At the null frequencies (multiples of 60 mHz), the noise sources are not suppressed. The other noise sources are coupled into the system when the arm is dominant. At very low frequencies, the cavity noise is dominant and follows the LISA requirement level.

\begin{figure}[H]
    \hspace*{-0.4cm}
    \includegraphics[width=7cm,height=10cm,angle=-90]{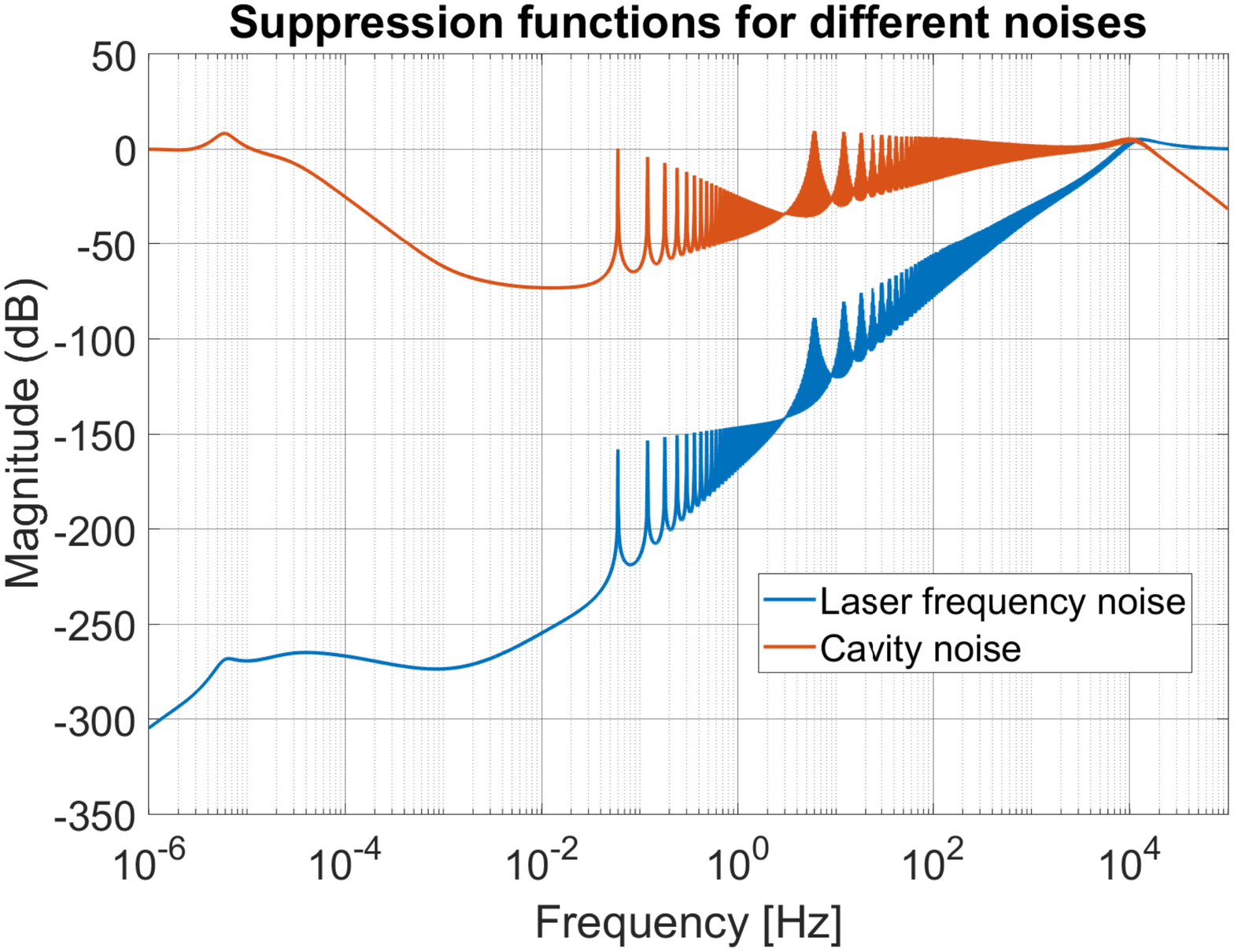}
    \caption{Suppression function of the system with respect to laser frequency noise and requirement-level cavity noise. The laser frequency noise is stabilised by both the arm and the cavity while the cavity noise suppression is provided by the arm in the LISA band}
    \label{Noise_suppression}
\end{figure}

The suppression function of the laser frequency noise, the first term in Equation~\ref{Totalnoise}, and the suppression function of the requirement-level cavity noise, the second term in the equation, is plotted in Figure~\ref{Noise_suppression}, with the controller discussed in Section~\ref{Controller_design}. This figure shows there is significant gain for laser frequency at all frequencies below 10 Hz and the suppression of cavity noise across the LISA science band.

\section{\label{Doppler_data}Doppler pulling and propagation in LISA} 
As the LISA constellation orbits around the Sun, the spacecraft follow independent orbits and the arm lengths are expected to change by upto 1\% ~\cite{Orbits}. The relative movement of the spacecraft cause Doppler shifts in the laser frequency given by $\nu_d$(t) = v(t)/$\lambda$, where v(t) is the relative velocity at which the spacecraft move and $\lambda$ is the wavelength of the laser (1064 nm). The maximum expected relative velocity between the spacecrafts is 5 ~ms$^{-1}$~\cite{LISAL3}, producing a one-way Doppler shift of approximately 5 MHz. As the lasers on the end station spacecrafts are phase locked to the laser on the primary spacecraft, the laser beam on returning back to the primary spacecraft will have twice the one-way Doppler shifts. The sum of the Doppler shifts will be taken when the phase is given to the common arm sensor. If the one-way Doppler shift of the laser travelling from spacecrafts \textit{i} to \textit{j} is $\nu_{D;ij}$, the effective Doppler shifts in the system can be shown as~\cite{mdual}
\begin{equation}
    \nu_{D,+}(t)= \nu_{D;12}(t) +\nu_{D;21}(t) + \nu_{D;13}(t) + \nu_{D;31}(t).
\end{equation}

 Based on orbital dynamics of LISA~\cite{Orbits,Orbits2}, we use a toy model that approximates the common Doppler shifts to be a sum of sinusoids with half-year and one year periods as shown below:
	\begin{equation}\label{Doppler_eqn}
    \nu_{D,+}(t)= \nu_1 \sin (\omega_1 t + \phi_1) + \nu_2 \sin(\omega_2 t +\phi_2).
	\end{equation}
 $\nu_1$ and $\nu_2$ are amplitudes of the two sinusoids of frequencies $\omega_1$ and $\omega_2$, along with phase shifts $\phi_1$ and $\phi_2$, respectively. This model can represent a smooth orbit evolution with (approximately) correct dynamics over the $\sim$1 month time frame for arm locking turn-on transient relevant to this paper. Pulling of the laser frequency in arm locking is a known issue~\cite{mdual,exp_mdual} that arises because the arm locking sensors have zero response at DC. On one hand, the cavity sensor can sense this Doppler pulling and allows correction at long timescales ( $> 10^4$ s). Conversely, the cavity sensor has finite linear range and limited gain in the LISA band, hence pulling must be minimised to ensure cavity lock is maintained. For this paper we consider this number to be $\pm20$~kHz, a fifth of the HWHM frequency (100~kHz).

An offset to the cavity lock point is undesirable as it introduces coupling from other noise sources, such as laser intensity noise. This section will demonstrate that, with the controller from Section~\ref{Controller_design}, laser frequency pulling reaches a maximum of 20 kHz at lock acquisition within the model framework. In laboratory testing using a  cavity and system with parameters similar to LISA, the residual laser frequency noise was seen to increase up to level of 100 Hz/$\sqrt{\textrm{Hz}}$ level with a 10 kHz offset off resonance on a PDH-stabilised laser with the setup in \cite{tiltlock}. Note: this noise would be still suppressed by the arm locking controller by the magnitude shown in Figure~\ref{Noise_suppression}.

\subsection{Lock Acquisition Timeline}\label{Timeline_section}
This section will focus on the potential lock acquisition timeline for LISA to employ arm locking while restricting the transient Doppler pulling caused by the feedback of the controller. For implementation of arm-locking, precise Doppler information is required for real-time subtraction from the phasemeter signal before feeding back to the laser. Section~\ref{Timescales} discusses two methods to measure the Doppler shift using only the hardware included in the baseline mission. The two techniques are 1) to use the inter-satellite range signal and average the beat note frequency for a sufficient time, and 2) use the inter-spacecraft absolute ranging system whose baseline deploys a Pseudo Random Code on the inter-satellite inter-spacecraft link~\cite{PRN_ref,PRN_ref2}. 
Depending on the noise performance of the cavity, either the cavity or the ranging readouts can be used for estimation. If the residual noise of the cavity is at requirements level, the estimates need at least 50000~s ($\sim$ 13.9~hours) of integration time to get the necessary accuracy. Instead, the absolute ranging system based on PRN ranging is relied on, requiring only 6300~s ($\sim$ 1.75~hours) of data to get a sufficiently accurate estimate of the Doppler trends. Cavity performance that approach the thermal limit~\cite{gracefo_cavity,cavitythermalnoise} will significantly reduce residual noise, allowing Doppler parameter estimates to converge within 1500~s ($\sim$ 0.5~hour) using the interferometer response only. These estimates (using either approach), can populate the phasemeter with a model estimate of the Doppler shifts based on Section~\ref{Lock_acquisition}.

\subsection{\label{Lock_acquisition}Arm Locking Acquisition}
\begin{figure}[H]
\hspace*{-0.6cm}
\includegraphics[width=7cm, height=10cm, angle=-90 ]{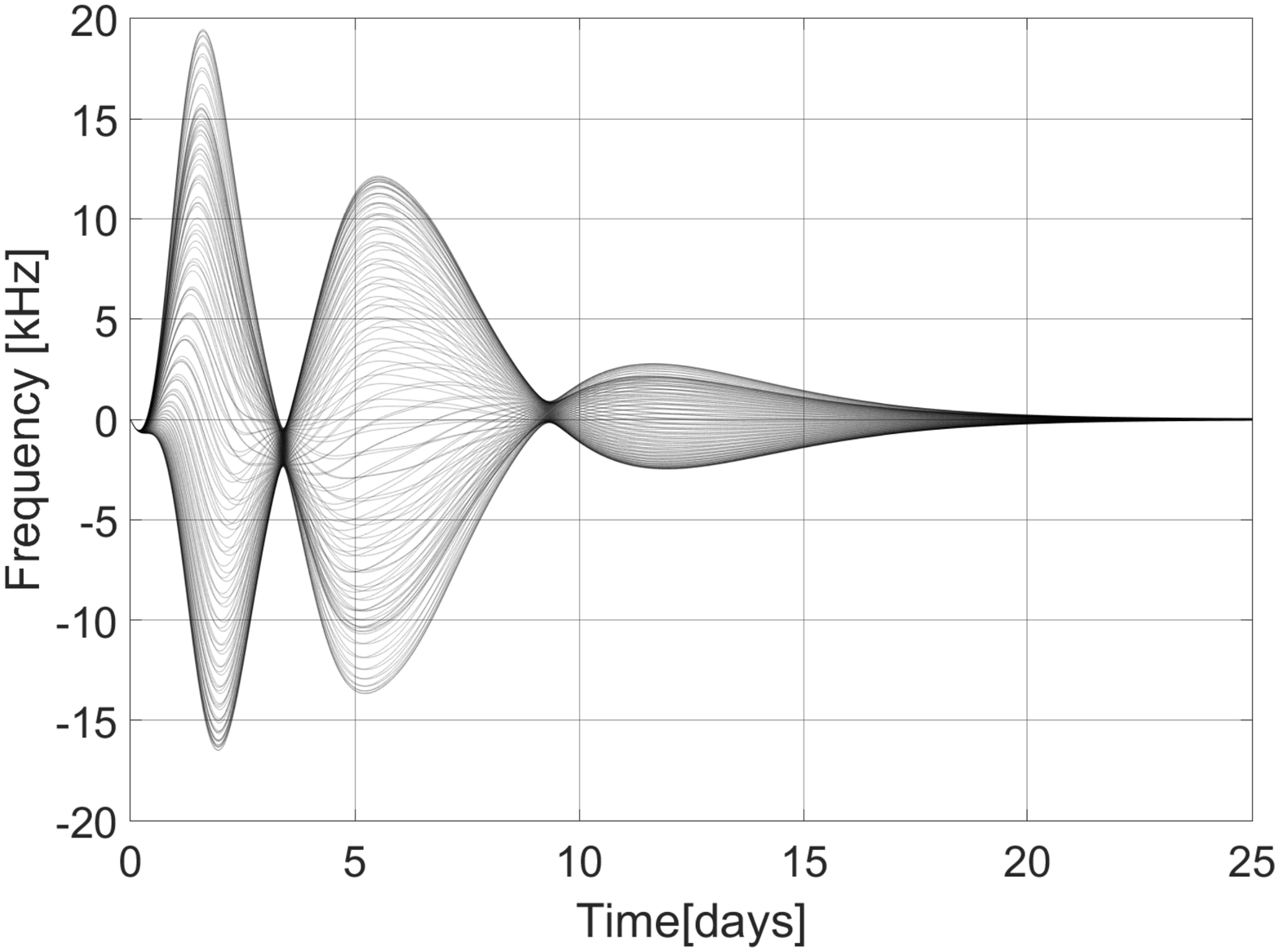}
\caption{Frequency deviation from cavity line center due to Doppler-error-pulling at lock acquisition. The estimate is calculated using the polynomial functions in Equation~\ref{Dopplerlock_poly}. The different traces corresponds to the lock acquisition Doppler pulling at different set points in the orbits with the error bounds in the model, given in Table~\ref{Dopplerreq}. The maximum Doppler pulling corresponds to the point in the orbits with the maximum Doppler rate.}
\label{pulling_orbits}
\end{figure}

   {When the arm locking controller is enabled on the primary spacecraft (Spacecraft 1) to stabilise the laser, estimates of the heterodyne frequency, including Doppler shifts, are subtracted from the phasemeter readings. Any error in Doppler shift estimate will represent a bias that cannot be suppressed by the arm locking controller. Accordingly, such a bias excites a transient response in the arm locking system and results in the frequency pulling behavior documented previously~\cite{mdual,exp_mdual}. Importantly, the introduction of the Fabry-Pérot (FP) cavity estimation provides necessary information to limit frequency pulling. This section describes the model for this transient behaviour. If the errors in Doppler shifts are small, then the transient laser frequency pulling can also be reduced to within the allowable limits of the cavity. The turn on transient of the controller can be represented by the closed-loop step response of the error between the actual and estimated Doppler shifts, and computed as 
    \begin{equation}\label{T7}
            \nu_{C} (t)=  L^{-1}\left( \frac{\left[\nu_{D;+}(s)-\left(\nu_{D;est}(s)\right)\right] V(s)}{s}\right),
      \end{equation}          

      \begin{equation}\label{T6a}
            V(s)=\frac{-G_1 (s)}{1+G_1 (s) P_{+} (s)+G_2(s) P_{pdh}(s)}.
\end{equation}}
$\nu_{D;est}(s)$ is an estimate of the common Doppler frequency. The term in square brackets consist of the error in estimation of Doppler shifts. $L^{-1}$ represents the inverse Laplace transform function of the system and V(s) is the transfer function of the Doppler shifts, with a similar analysis in Section~\ref{Noise_prop}. \begin{table*}
\centering
\caption{Parameter requirements for orbital knowledge in order to meet lock acquisition conditions using the polynomial model with the controller shown in Fig 14. Each parameter's error limit is checked in combination with the errors of other parameters in Monte Carlo simulations. The achievable levels are cross-checked with estimation using Fabry-Pérot (FP) cavity estimation in~\cite{mdual} and PRN ranging, or Thermal Noise limited (TNL) cavity estimation, as described in Section~\ref{Timescales}.}    
\label{Dopplerreq}      
\begin{tabular}{|c|c|c|c|c|}     
\hline       
  Parameter  &  Actual/Max  & Max error & Fractional & Estimation\\ & Value  & tolerance ($\pm$) &change  &methods \\ 
\hline  
$\nu_0$ & 12 MHz & 10 Hz  & 8.33 x $10^{-7}$  & FP cavity estimation/ TNL cavity estimation\\
$\gamma_0$ &   4 Hz/s & 60 $\mu$Hz/s  & 1.5 x $10^{-4}$ & PRN ranging/ TNL cavity estimation\\
$\alpha_0$ & -1.2 $\mu$Hz/s$^2$ & 5 nHz/s$^2$  & 4.17 x $10^{-3}$& PRN ranging/ TNL cavity estimation\\ 
\hline                  
\end{tabular}
\end{table*}
A simplification of the Doppler shifts can be done for smaller timescales, where the Doppler shifts at that instant can be modelled as a second order polynomial equation shown below~\cite{mdual}:

\begin{equation}\label{Dopplerlock_poly}
    \nu_{D;est} (t)=  \left(\nu_{0;+}+ \gamma_{0;+}t + \frac{\alpha_{0;+}t^2}{2}\right),
\end{equation}
        
where $\nu_{0;+}$, $\gamma_{0;+}$ and $\alpha_{0;+}$ are the estimates of the Doppler shift, the first derivative of Doppler shift (Doppler rate) and second derivative of Doppler shift (Doppler acceleration) at the instant when the controller is just turned on. The error limits, based on the results of Fabry-Pérot cavity estimation for 200~s given in~\cite{mdual},were sufficient due to the wide limits afforded for a free running laser (upto 10 MHz). For this work, we require that the worst case Doppler pulling be bounded by $\pm20$ kHz. \newline\newline
The toy model values of $\phi_1$ and $\phi_2$ in Equation~\ref{Doppler_eqn} determine the point in the orbit paths at which laser lock is acquired. At the point of higher Doppler rate, we expect the model to have more stringent error requirements compared to locking at points where the Doppler rate have relatively lower amplitude. Figure~\ref{pulling_orbits} show the Doppler pulling when locking at different orbital points using a polynomial model with the error bounds given in Table~\ref{Dopplerreq}. The traces are shown to be restricted within $\pm$20 kHz and thus provide us the flexibility to engage the arm locking controller at any point in the spacecraft orbit, without causing any significant Doppler pulling. The poor model fit (due to the difference in Equation~\ref{Doppler_eqn} and~\ref{Dopplerlock_poly}) results in an intrinsic ‘un-modelled’ Doppler error propagated into the control system. Even with ideal parameter knowledge, the polynomial model is incapable of fully resolving Doppler evolution and will result in a Doppler pulling of up to 17 kHz (relative to the assumed sinusoidal orbit description in Equation~\ref{Doppler_eqn}).

Adaptation of a more complex Doppler model can theoretically eliminate this source of systematic estimation error and associated intrinsic Doppler pulling component. However, the complexity and the number of variables that need to be estimated requires further analysis. A preliminary analysis of the sinusoidal model (along with the error bounds) is given in Appendix B, based on the toy model in Equation~\ref{Doppler_eqn}.

\begin{figure}[H]
\hspace*{-0.4cm}
\includegraphics[width=7cm, height=10cm, angle=-90 ]{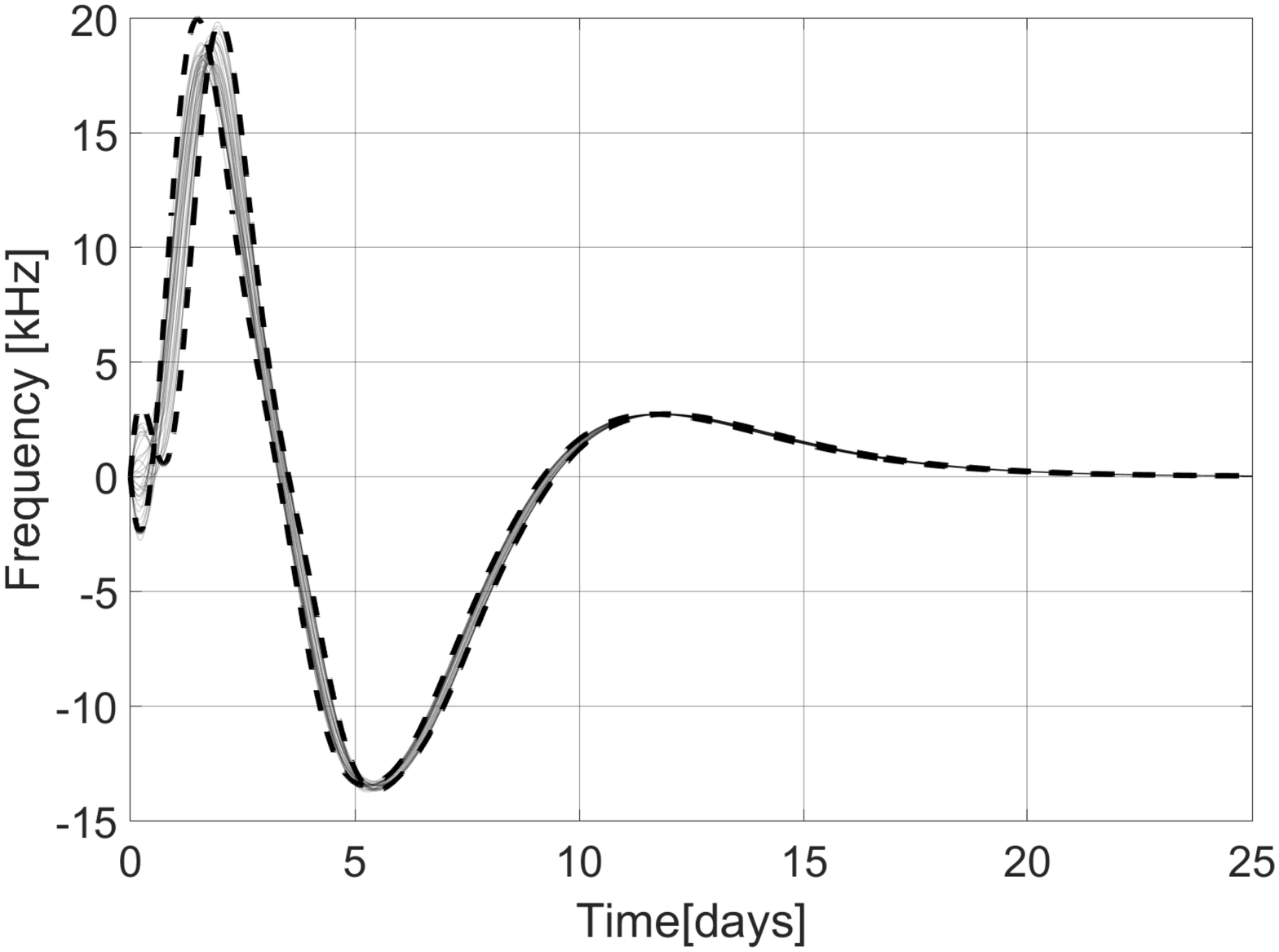}
\caption{Monte Carlo simulations of frequency deviation from cavity line center due to Doppler-error-pulling at lock acquisition. The estimate is calculated using the polynomial functions in Equation~\ref{Dopplerlock_poly}. The black traces shows the various Monte Carlo simulations with the dashed traces showing the pulling with the maximum error limits, given in the Table~\ref{Dopplerreq}. The maximum pulling is limited to 20 kHz within the line-width, not breaking the resonance of the cavity.}
\label{pulling_transient_poly}
\end{figure}

Further investigations were done to test the error bounds of the Doppler parameter estimates by running Monte-Carlo simulation (100 runs). With the orbital set-point selected to give the maximum pulling, Figure~\ref{pulling_transient_poly}
shows that the polynomial model and associated derived error bounds in Table~\ref{Dopplerreq} can restrict laser frequency pulling to within the $\pm$20 kHz requirement. After 25 days, the pulling would reach steady state that is limited by $\pm27$~Hz.

During the steady state period, we expect the spacecraft to undergo orbits corrections or engage spacecraft thrusters, which would come as a step or ramp function in the Doppler error. Analysis on this disturbance in the steady state of the Doppler pulling indicate an excitation of a transient response in the control loop comparable to initial lock acquisition. If the frequency deviations are within the requirements shown in Table~\ref{Dopplerreq} for $\nu_0$ and $\gamma_0$, the laser will maintain lock with the cavity.

\section{\label{Timescales}Doppler Frequency Estimation}
The previous section presented details of the Doppler knowledge requirements at controller activation. This section presents the convergence time required to estimate the Doppler shifts with sufficient accuracy. The methods presented here require nominal LISA measurements, without arm locking engaged, for a few thousand seconds to allow Doppler shift estimate of sufficient accuracy. Once sufficient Doppler shift knowledge is achieved, the arm locking controller can be activated, and the associated improvement in laser frequency noise suppression achieved. The two methods of measurement considered are: 1) LISA phasemeter measurement containing the nominal interferometer response using cavity-stabilised laser, 2) using the LISA baseline inter-spacecraft range measurement~\cite{PRN_ref,PRN_ref2}. The required estimation times are determined based upon the observation time over which the weighted Allan variance of the residual noise reduces to below the required RMS level.

The PRN ranging system uses a Pseudo Random Binary Sequence (PRBS) modulation to time stamp the outgoing light with a clock-like signal as it leaves the spacecraft, providing a reference with which the clocks can be aligned and the distance can be measured between the spacecrafts. This is limited by PRBS code noise and shot noise, with a residual displacement noise expected of order $\sim$0.1 m/$\sqrt{\textrm{Hz}}$~\cite{PRN_ref,PRN_ref2}.

The FP cavity estimation method described in Appendix A of~\cite{mdual}, has a residual noise derived from the cavity noise coupled through an open arm sensor. For this calculation, we make two estimates of Doppler error based on two levels of cavity residual errors: 1) the cavity residual given in Equation~\ref{N3}, which is a conservative estimate, 2) the cavity residual such as in GRACE-FO~\cite{gracefo_cavity} or in~\cite{cavitythermalnoise}, that approach the cavity thermal noise limit (TNL)~\cite{thermallimit}, the best possible case (in reality, the cavity will likely be between these two bounds), modelled as
\begin{equation}\label{thermal_limit}
    \nu_{cavity} \approx \nu_{thermal\_limit}= \frac{0.1}{\sqrt{f}} \hspace{0.3cm} \textrm{Hz}/\sqrt{\textrm{Hz}}.
\end{equation}
The effective residual displacement noise can be derived as:
\begin{equation}
    \begin{split}
     \Delta x_{residual}(s)  & = \frac{\Delta\phi_{cavity}(s) *P_+(s) * \lambda}{2} \\ & = \frac{ \nu_{cavity}(s) * P_+(s) * \lambda}{2s}.
     \end{split}
 \end{equation}
The estimator performance is predicted based upon the expected weighted Allan deviation over time. The Allan deviation was derived from the time domain using numerical simulation of the relevant noise spectra. To mitigate the effects of temporal leakage, we apply a windowed and overlapped Allan variance estimation from~\cite{window_Allan}. The following equation captures the operation of windowing in the overlapped Allan variance formula,
 \begin{figure}[H]
    \hspace*{-0.6cm}
     \includegraphics[width=6.75cm, height=9.75cm, angle=-90]{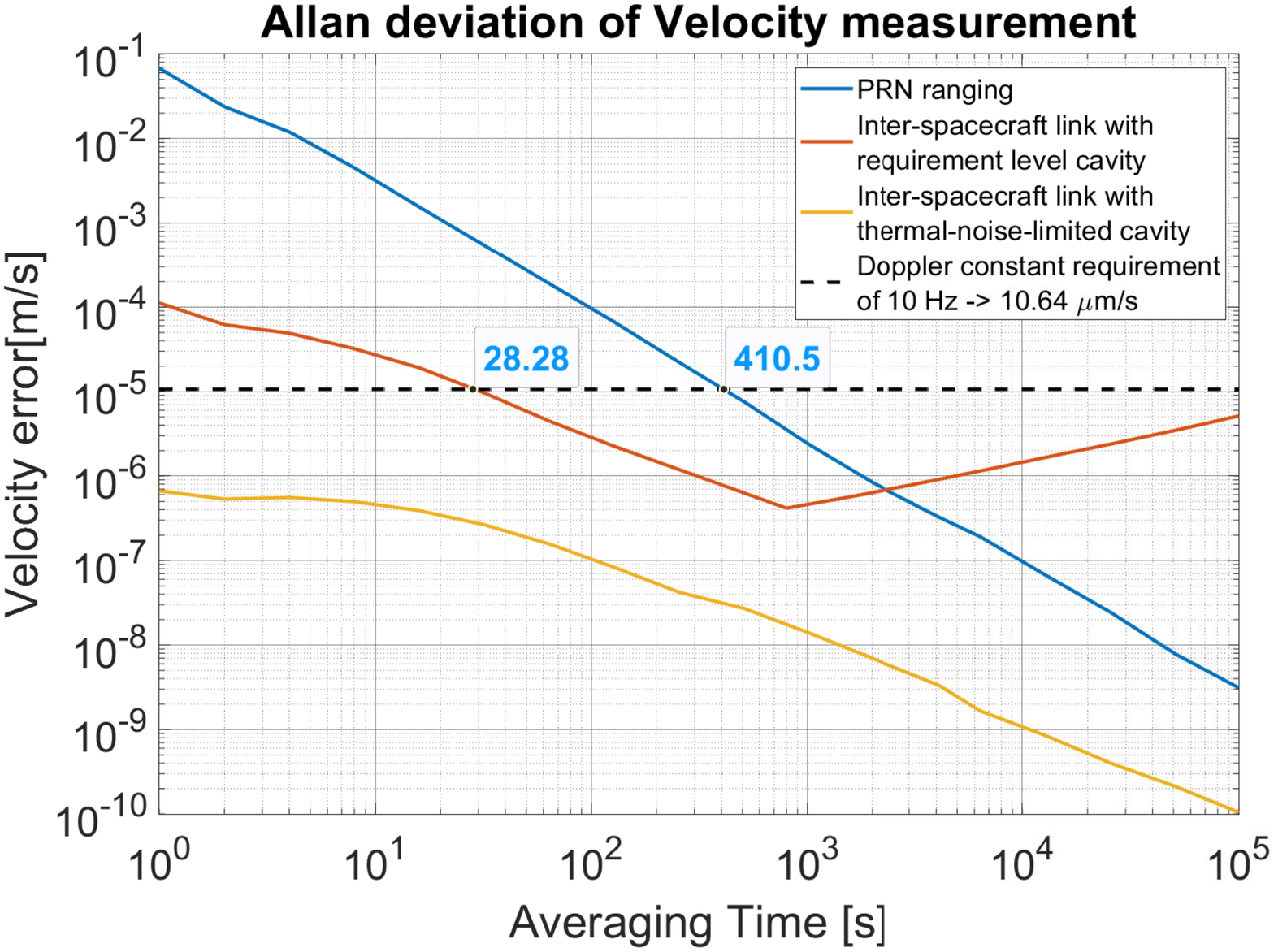}
    \caption{Allan deviation for the velocity measurement between the spacecrafts. The dashed line indicates a Doppler constant requirement of 10 Hz, corresponding to 10.64 $\mu$m/s. The PRN ranging scheme has a slope of -3/2 while the requirement-level cavity-arm sensor has a slope of -1/2 till around 500~s and then rolls up with slope of 1/2. Thermal-limited cavity-arm sensor has a slope of -1 after 8.335~s. The time required for the deviation to reach the error limit is 411~s for PRN while the cavity-sensor can reach it under 30~s and thermal-limited cavity sensor can always realise this requirement (within 1~s).}
    \label{Allan_Doppler_constant}
\end{figure}
\begin{figure}[H]
    \hspace*{-0.6cm}
    \includegraphics[width=7.25cm, height=10.25cm, angle=-90]{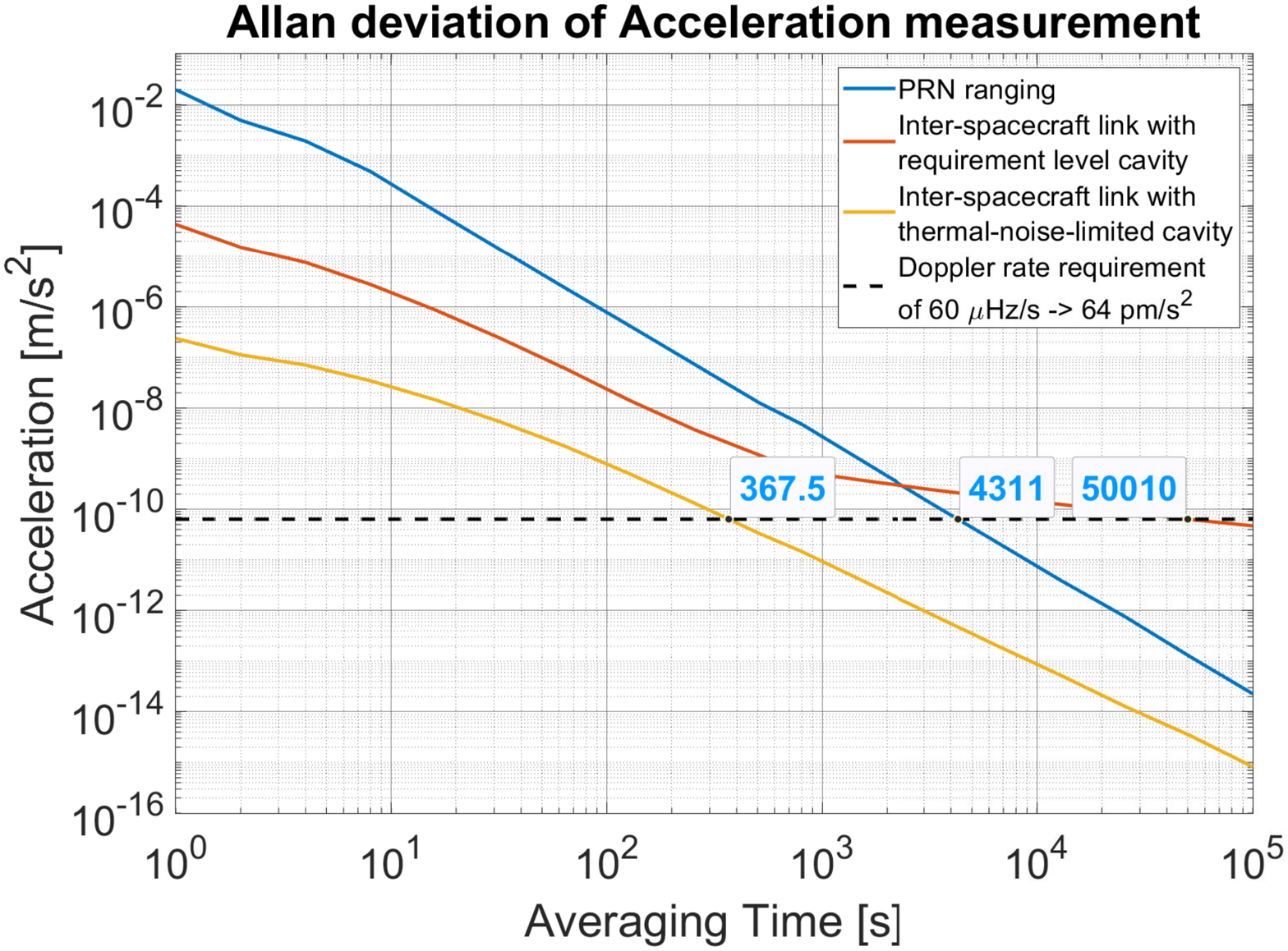}
    \caption{Allan deviation for the acceleration measurement between the spacecrafts. The dashed line indicates a Doppler rate requirement of 60 $\mu$Hz/s, corresponding to 64 pm/s$^2$. The PRN ranging scheme has a slope of -5/2 while the requirement-level cavity-arm sensor has a slope of -3/2 till 500~s and then rolls off with -1/2. The thermal-limited cavity-arm sensor has a slope of -1 till 8.335~s and rolls off with -2. The time required for the deviation to reach the error limit of 60 $\mu$Hz/s is 4311~s for PRN while the cavity-sensor requires 50010~s. The thermal-limited cavity sensor can estimate the Doppler rate to the required level within 370~s.}
   \label{Allan_Doppler_rate}
\end{figure}

 \begin{figure}[H]
    \hspace*{-0.65cm}
    \includegraphics[width=7.27cm, height=10.25cm, angle=-90]{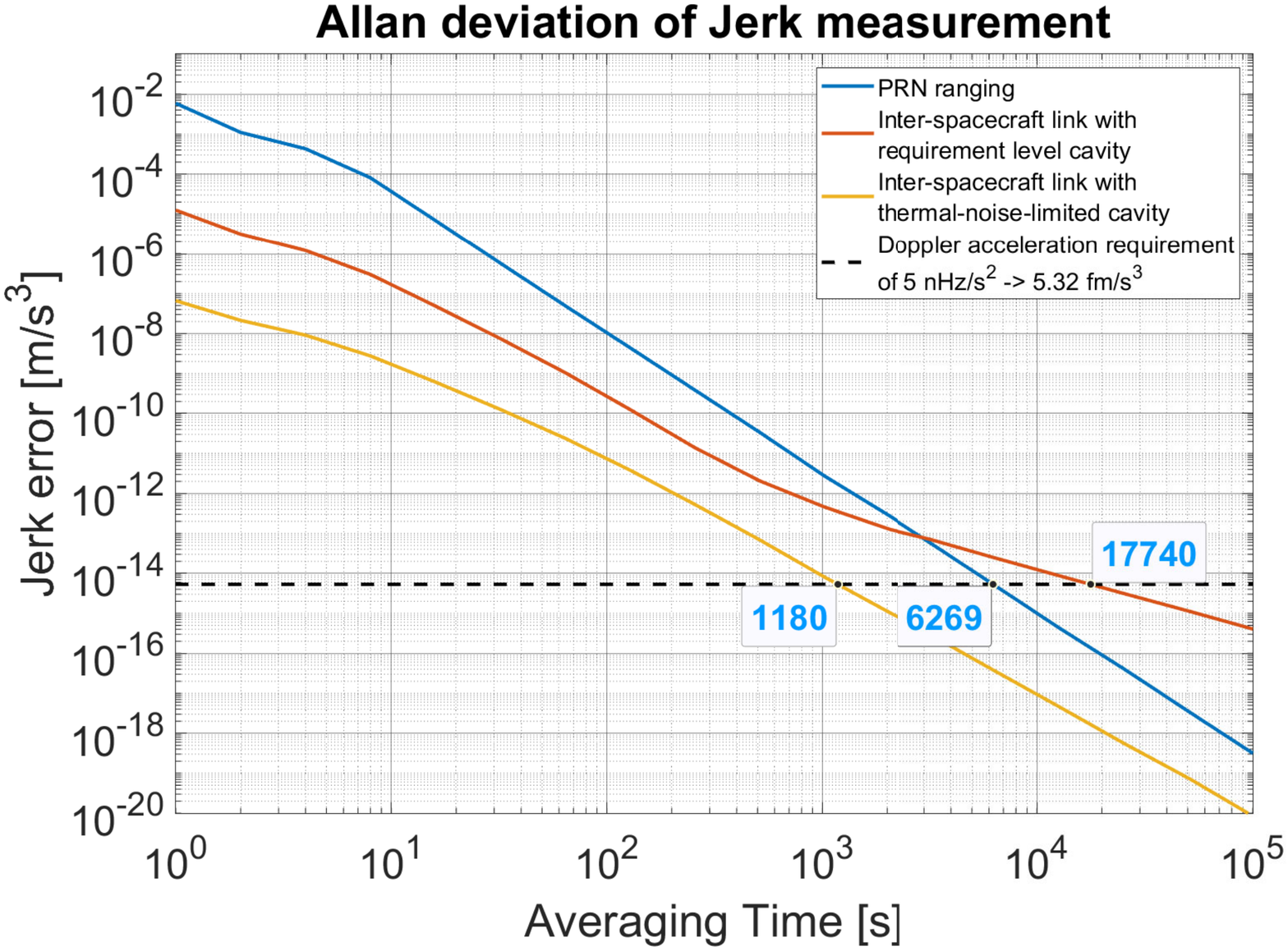}
    \caption{Allan deviation for the jerk measurement between the spacecrafts.  The dashed line indicates a Doppler acceleration requirement of 5 nHz/s$^2$, corresponding to 5.32 fm/s$^3$. The PRN ranging scheme has a slope of -7/2 while the requirement-level cavity-arm sensor has a slope of -5/2 till 500~s and then rolls off with -3/2. The thermal-limited cavity-arm sensor has a slope of -2 till 8.335~s and rolls off with -3. The time required for the deviation to reach the error limit is 6270~s for PRN while the cavity-sensor requires 17740~s. The thermal-limited cavity sensor can estimate the parameter to the required level within 1200~s.}
    \label{Allan_Doppler_acceleration}
\end{figure}
\vspace{-0.5cm}
\begin{equation}
    \sigma^2_y (\tau)=\frac{\sum\limits_{j=1}^{M-3m+2}\{\sum\limits_{i=j}^{j+m-1}\left(w_m(i)\left[y(i+m)-y(i)\right]\right)\}^2}{2m^2(M-2m+1)},
\end{equation}

where m=$\tau/\tau_0$, $\tau_0$ is the sampling rate of the measurement data, y,of length, M. $w_{m}$ is the window function of length, m, corresponding to each $\tau$, the averaging time. The window function used in this paper is the normalised Blackman-Harris filter giving the correct Allan deviation slopes for the different higher noise slopes. Figures~\ref{Allan_Doppler_constant}, \ref{Allan_Doppler_rate} and~\ref{Allan_Doppler_acceleration} showcase the Allan deviation (after normalisation) for each parameter (Doppler constant, rate and acceleration) using the residual displacement sensitivity as above in time domain. The data points highlighted in each figure are the minimum times that is required for the noise to be averaged to meet the error limits in Table~\ref{Dopplerreq}.

\begin{table}[H]
\caption{Minimum averaging time required using different estimation methods to reduce the Doppler terms to the error limits in Table~\ref{Dopplerreq}. }
\centering
\begin{tabular}{m{1.77cm} m{1.78cm} m{1.78cm} m{1.33cm} m{1.3cm}}     
\hline\hline       
Doppler & \multicolumn{2}{c}{Inter-spacecraft link} & PRN & Required 
\\ \cline{2-3}term & Thermal & Requirement & absolute & precision  
\\  & limited & level & ranging & (Doppler)
\\  & Cavity & Cavity & & trends)\\
\hline      
Constant &  1~s &  30~s & 411~s & 10 Hz \\
Rate &  370~s &  50010~s & 4311~s & 60 $\mu$Hz/s\\
Acceleration & 1200~s &  17740~s & 6270~s & 5 nHz/s$^2$\\

\hline                  
\end{tabular}
\label{Estimation_Timescale}
\end{table}
From the Figures~\ref{Allan_Doppler_constant}, \ref{Allan_Doppler_rate} and~\ref{Allan_Doppler_acceleration}, it can be observed for that the thermal-noise-limited cavity-arm sensor (yellow trace) can be used for estimating all the parameters within a short timescale ($<$ 0.5 hour). The PRN (blue trace) would perform better than the cavity, if the residual noise is near the LISA pre-stabilisation limit (red trace).

If the cavity does not reach the thermal-noise-limit, the cavity can still be employed for determining the Doppler constant parameter, while the PRN can be used for determining the higher derivatives of the Doppler shifts (which necessitate longer integration times for estimation). The requirement-level cavity-arm sensor is limited by the random walk function at lower frequencies, and after 500~s has more residual noise. This estimation needs to be employed every time the laser acquires lock with the spacecrafts and can be used only in the beginning of the laser lock for estimating the constant Doppler shift. The PRN technique would be ideal for determining the other Doppler parameters due to the residual being white noise and hence over time, the measurement errors reduce over time, by the square root of time, for velocity, and more quickly for acceleration and jerk.

\section{\label{Controller_design}Controller Design}
{
From the descriptions, transfer functions, and the effects of noises on the output, and the LISA goals, the main requirements for the controllers can be derived. However additional requirements were also included to  make the controller more robust and perform better during the mission lifetime.
\begin{enumerate}
    \item{The phase margin at unity gain crossings must be more than $30\degree$ (open loop phase within $\pm150\degree$). Additionally, the same condition is applied for cross-over frequencies between arm and cavity.}
    
	\item{The arm locking control system is to have at least 15 times the gain of the cavity control system at $10^{-4}$~Hz and at least 100 times the FP cavity gain at 1 Hz. These are soft conditions to prioritise the bandwidth of the arm locking sensor within the LISA science band.}

\end{enumerate}
Keeping these conditions in mind, the controller design can be split in two parts.
\subsection*{\label{Controller1}Controller 1: Arm Locking Controller}

This controller is used with the arm locking sensor, and is used mainly to control bandwidth of the Arm sensor, shown as G$_1$ in Figure~\ref{Model_n}. This controller can be further split in three stages as follows:
\newline

Stage I: Stage I is primarily a controller with a fractional slope of 2.3 and a unity gain frequency at 13.63 kHz. The integrator scales the sensor information and provides feedback back to the laser source at lower frequencies. This provides the necessary suppression below LISA requirements and the slope ensures that the cavity will dominate the response at higher frequencies. The implementation of the controller is done using a sum cascade of low-pass filters as shown in Table~\ref{LP_table} in Appendix B.
\newline

Stage II: Stage II consists of a cascade of 7 high pass filters in the low frequency band ($ < $ 10 $\mu$Hz), to ensure the cavity dominates at lower frequencies, allowing a reduction of the Doppler pulling induced by the arm sensor. The different high-pass filters allow for a smooth transition of the arm and the cavity in terms of the phase stability of the system. The combined response gives 7 x 20 db = 140 db/decade of suppression of Doppler shifts at very low frequencies ($< 0.1 \ \mu$Hz). Implementation and optimization of such a filter structure is a topic for future work, with a particularly focus required upon stability and the partitioning of the controller between software and firmware.

\begin{figure}[H]
    \hspace*{-0.5cm}
    \includegraphics[width=7cm,height=10.25cm, angle=-90]{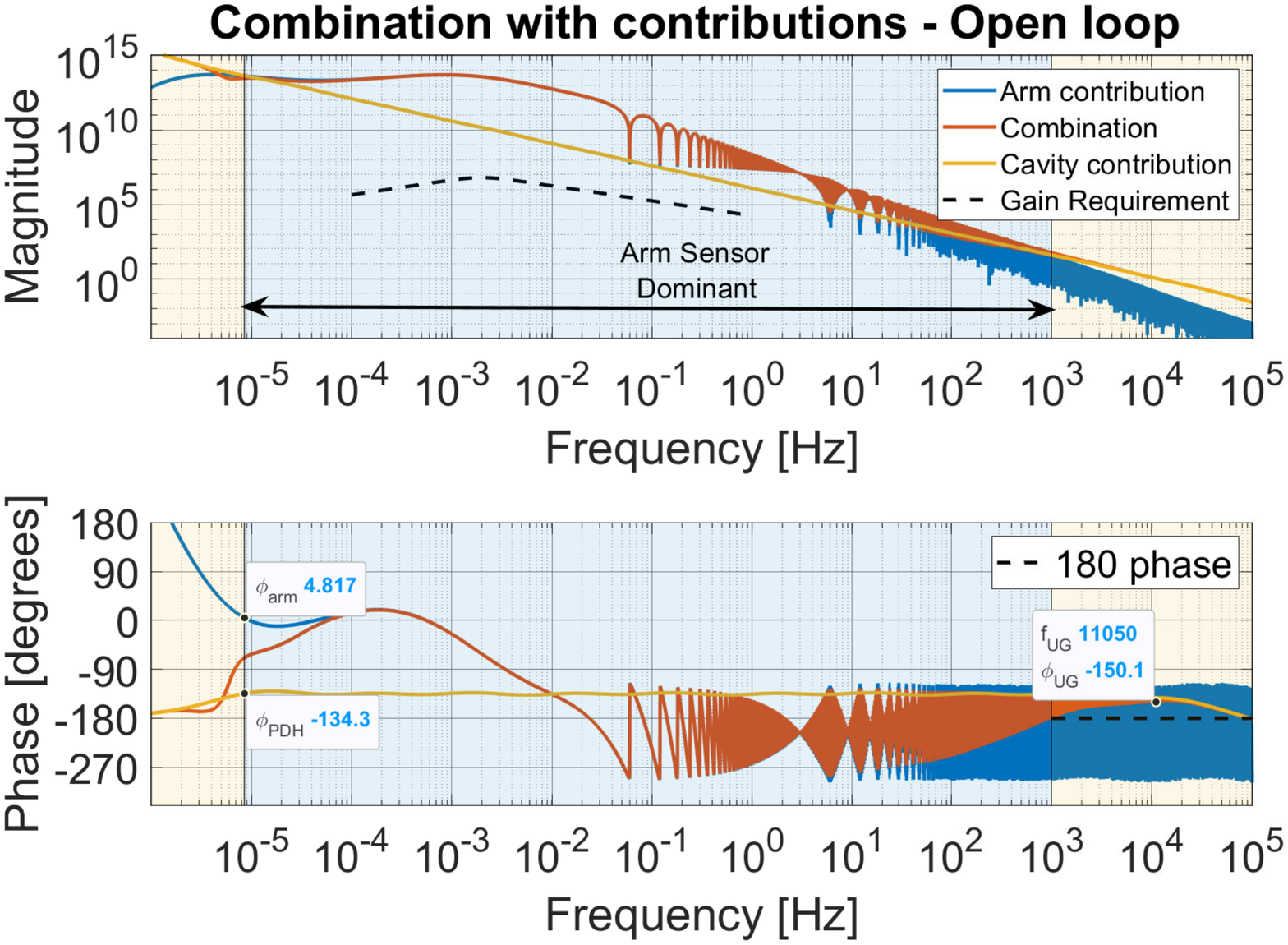}
    \caption{Open loop gain of the controller. The UGF is at around 11 kHz and the phase margin is within 30$\degree$. The arm sensor gain is dominant from $\sim$10 $\mu$Hz to 1 kHz, while the cavity is dominant outside this region. The blended sensor must maintain sufficient phase margin where the cavity and arm-locking gain are equal to prevent noise amplification, such as the approximately 40$\degree$ achieved at the low cross-over point ($\sim$10 $\mu$Hz). Phase margin of $\sim20\degree$ is obtained around the nulls of the arm-locking sensor, where the arm-locking gain crosses the cavity gain curve, resulting in an increased sensitivity to cavity noise in this region.} 
    \label{fig:open_loop}
\end{figure}
Stage III: Stage III is a lag compensator to decrease phase of the arm controller and thus allow the phase margin at the lower cross-over frequency between the arm and the cavity to be within the requirements.\newline\newline

The combined response for Controller 1 is shown below with the values in Table~\ref{Controller_table}:
\begin{equation}\label{G1}
\begin{split}
   G_1(s) =&\left(\frac{g_1}{s}\right)^{2.3}  \left(\frac{s}{s+p_{h_1}}\right)^{5}
    \left(\frac{s}{s+p_{h_2}}\right)^{2}
    \\
    &\left(g_{lc}\left(\frac{s+z_{lc}}{s+p_{lc}}\right)\right).
\end{split}
\end{equation}

\subsection*{\label{Controller2}Controller 2: Cavity Controller}
{This controller is used along with the cavity sensor, as shown as G$_2$ in Figure~\ref{Model_n}. Within the LISA band, the controller gain follows a fractional f$^{-3/2}$ slope with a UGF of 7.32 kHz. The positive slope can make sure that the cavity does not dominate arm locking within the LISA band but dominate at higher frequencies, while at lower frequencies the cavity will have a larger gain due to the high pass filters in Controller 1.
\begin{equation}\label{G2}
    G_2(s)=\left(\frac{g_2}{s}\right)^{1.5}.
\end{equation}

The implementation of the controller is done using a sum cascade of low-pass filters as
shown in Table~\ref{LP_table_2} in Appendix B.
}
\begin{table}[h]
\centering
\caption{Values for the gain, poles and zeros for the controllers as shown in Equations~\ref{G1} and~\ref{G2}}             
\label{Controller_table}      
\begin{tabular}{c c}    
\hline\hline       
Parameter & Value \\ 
\hline                    
   $g_1$ & 2$\pi$ x 1.36 x $10^4$\\
   $g_2$ & 2$\pi$ x 7.32  x $10^3$ \\
   $p_{h_1}$ & 2$\pi$ x 1.29x$10^{-6}$ rad/s\\
   $p_{h_2}$ & 2$\pi$ x 1.16 x $10^{-3}$ rad/s\\
   $z_{lc}$ & 2$\pi$ x $10^{-4}$ rad/s\\
   $p_{lc}$ & 2$\pi$ x 4.5 x $10^{-6}$ rad/s\\
   $g_{lc}$ & 0.045\\
\hline                  
\end{tabular}
\end{table}

\begin{figure*}
\includegraphics[width=18cm]{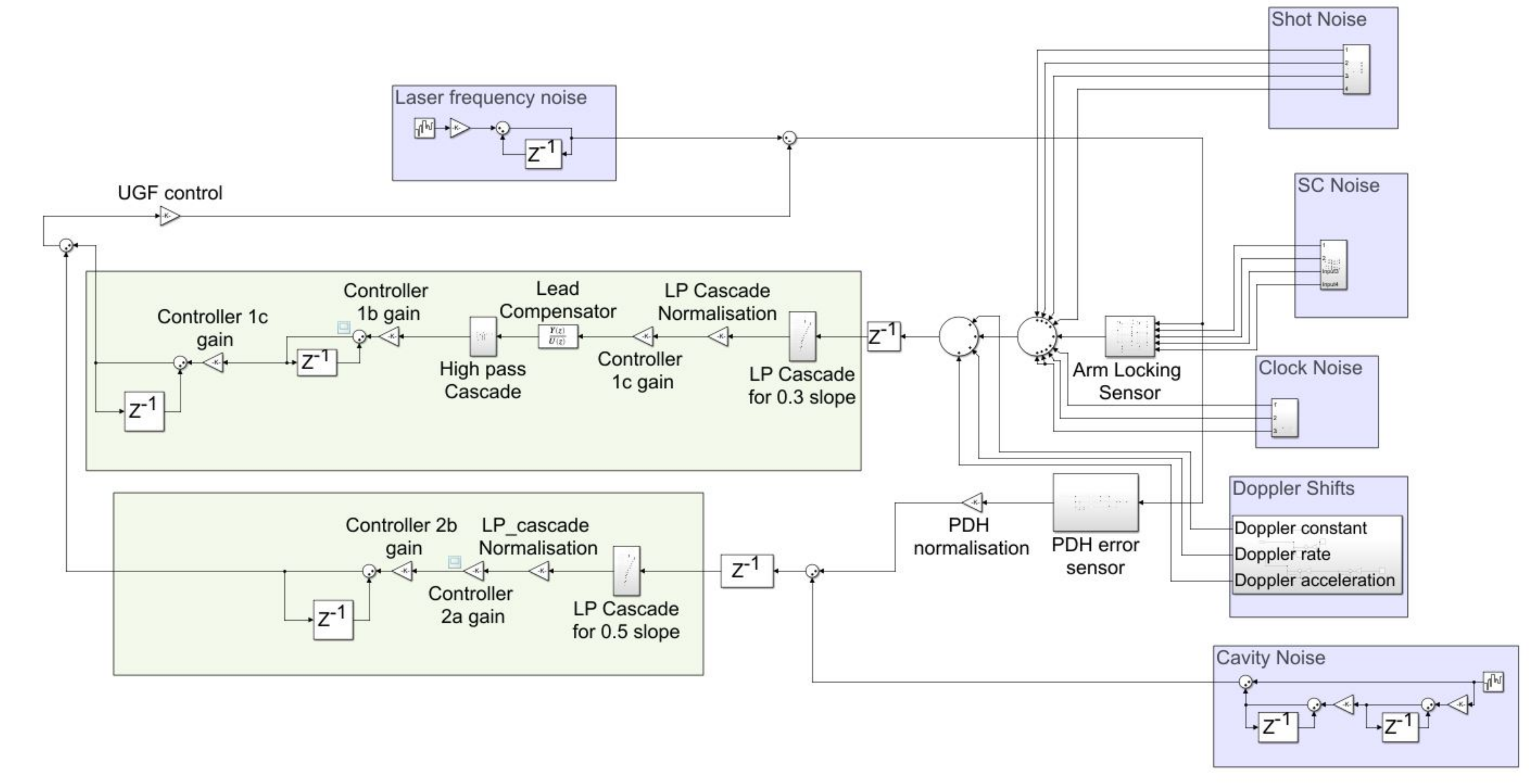}
     \caption{Simulink Model illustrating the combination of arm and PDH locking. The purple boxes contain the noise sources considered in this model while the green boxes contain the controllers for each of the sensor. The UGF of the combined controllers is 500 Hz, with the sampling frequency at 10 kHz.}
     \label{Sim_Model}
 \end{figure*}

\section{\label{Simulink_model}Simulink Model}
Time domain simulation for a representative control system was performed for model validation. Due to computational limitations, simulations of the long (1 month) Doppler transients with fast control loops (UGF at 10 kHz) was prohibitive. Instead, a simplified problem was studied to validate the modelling procedure utilized for LISA predictions. The proposed controller architecture was tested using a discretized Simulink time-domain simulation shown in Figure  \ref{Sim_Model}. In the analysis, the round trip time is taken to be 1~s with the controller unity gain frequency scaled down to 500 Hz. Results are obtained by running a discrete solver for fixed step size at the sample rate of 10 kHz.

The challenge of computationally exhaustive algebraic loop at higher sampling frequencies was solved by added an explicit pipeline delay, converging the loop within a time step, an accommodation needed to run the simulation. 
This posed another problem as the phase introduced by this delay led to instability of the controller at unity gain frequency. The additional phase introduced by the delay can be measured as
\begin{equation}
    \phi_{\textrm{delay}}(f)=2\pi f\tau_{\textrm{delay}} ,
\end{equation}
where $\tau_{\textrm{delay}}=1/f_{\textrm{sampling}}$, the sampling frequency. To ensure the delay's contribution does not affect model conclusions, the sampling rate should be at least a factor of 20 higher than the unity gain frequency. Hence for UGF of 10 kHz, a sampling frequency of 200 kHz was required. This will allow at most only $18\degree$ of phase variation at the UGF and hence be less prone to instability as the phase margin is $30\degree$. For the purpose of simulation and evaluation, instead of having a higher sampling frequency, the gains of the controllers were scaled down using a overall scale factor to get a lower unity gain frequency. Thus, the final model was designed with controller of unity gain frequency at 500 Hz with a sampling frequency of 10 kHz. In this modified control system, the arm is dominant from 0.2 mHz till 200 Hz, while the cavity is dominant in the remaining frequency band.

The Doppler shifts were added to the system as errors in the different Doppler trends similar to the analysis in~\cite{mdual}. The resultant pulling will be increased by the shorter round trip and at the same time, reduced due to a lower arm controller bandwidth. The Simulink data is compared with the predictive Doppler pulling model used in Section~\ref{Doppler_data} using the modified control system. 

 \begin{figure}[H]
    \hspace*{-0.6cm}
    \includegraphics[width=7.25cm, height=10.25cm, angle=-90]{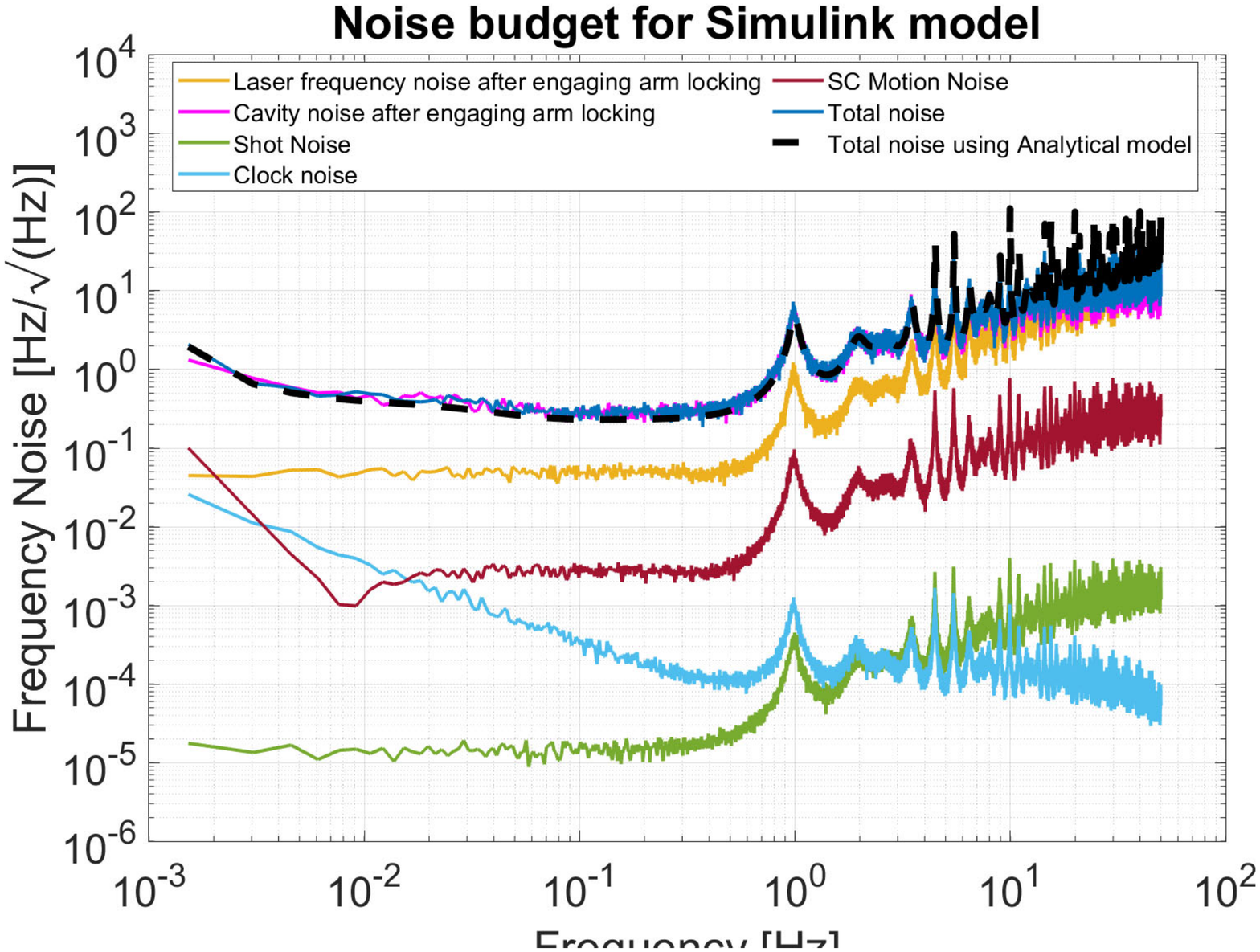}
    \caption{Noise  Budget  using  the same noise models and transfer functions in Section~\ref{Noise_prop}. The controller has a reduced unity gain frequency at 500 Hz and the sampling frequency for this simulation is 10 kHz and has been decimated to 100 Hz. Similar to Figure~\ref{fig:noise_budget}, the main contributing  noise  source  is  the requirement-level cavity  noise (pink trace) with the arm arm controller engaged at lower frequencies.}
    \label{Simulink_NS}
\end{figure}
Figure~\ref{Simulink_NS} shows the noise spectrum using the time-domain data from the Simulink model. Similar to the analytical noise budget in Figure~\ref{fig:noise_budget}, the main contribution is given by the requirement-level cavity noise that is suppressed after engaging the arm controller. The other noise sources couple in a similar way to in Figure~\ref{fig:noise_budget}.  Figure~\ref{Simulink_Doppler} shows the Doppler pulling of the scaled system with the associated analytical model. The close agreement between analytic and simulated noise models provide validation of the analytical modelling approach in Sections~\ref{Model} and~\ref{Doppler_data}.

\begin{figure}[H]
    \hspace*{-0.6cm}
    \includegraphics[width=7.25cm, height=10.25cm, angle=-90]{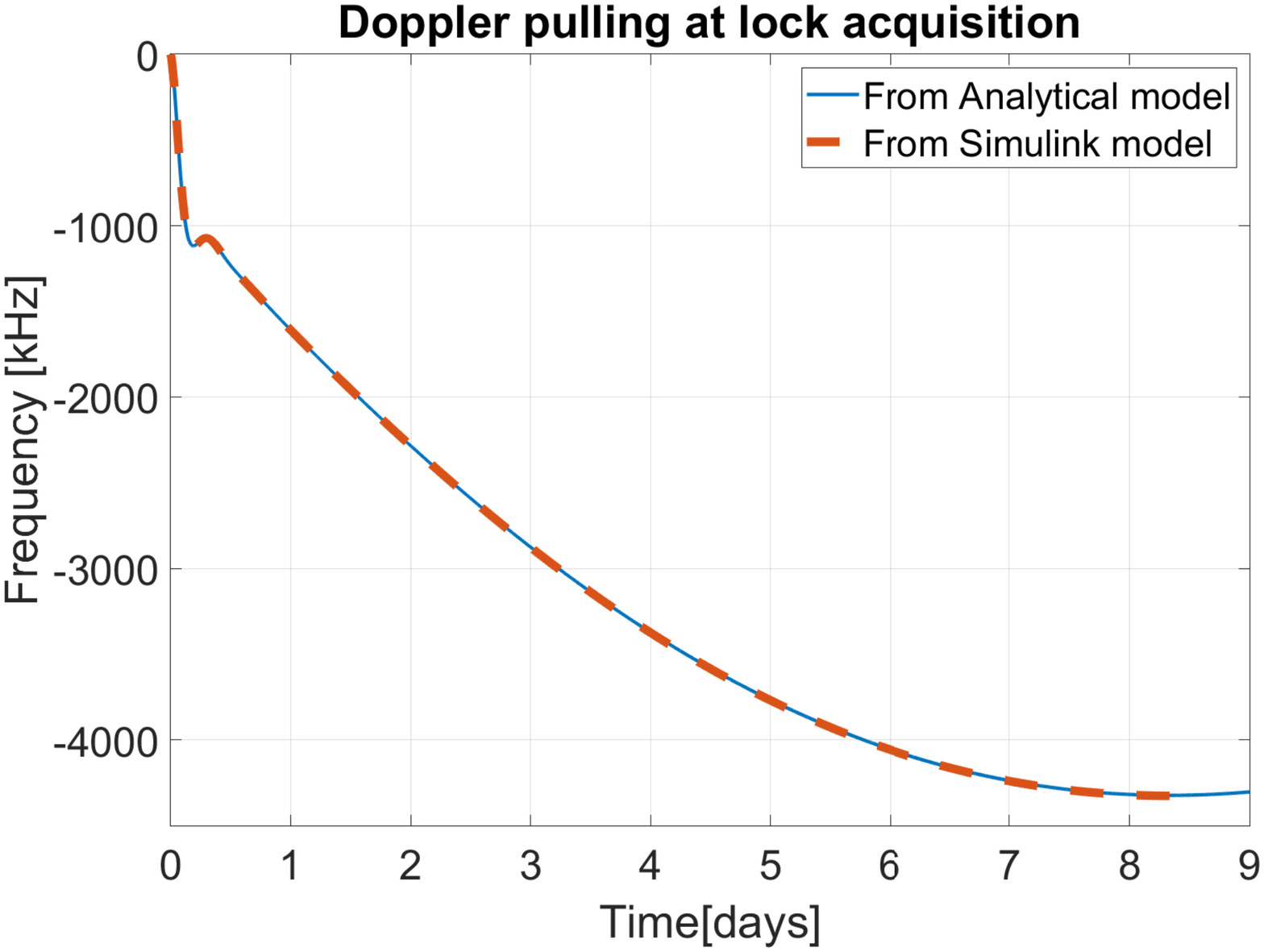}
    \caption{Doppler pulling of the system with an errors in the Doppler terms as $\nu_0$ = 1.682491 Hz, $\gamma_0$ = 0.084615 Hz/s and $\alpha_0$ = - 0.37239 nHz/s$^2$.  The controller has a reduced unity gain frequency at 500 Hz and the sampling frequency for this simulation is 10 kHz and has been decimated to 100 Hz. The predictive model is able to match exactly the system performance from Simulink}
    \label{Simulink_Doppler}
\end{figure} 
\section{\label{Discussion}Discussions}
This paper has proposed a hybrid arm- and cavity- locking scheme for LISA. The demonstrated highlights are;
\begin{enumerate}
      
  \item The hybrid control system increases the noise suppression by 3 orders of magnitude over most of the LISA science band (a factor of 4500 at 10 mHz). This large suppression can be considered as risk-reduction against TDI, which continues to be one of LISA’s main areas of active research, either by providing margin over baseline second-generation TDI or allowing the deployments of simpler first-generation scheme. The noise reduction can increase the margin for the baseline second-generation TDI or be sufficient to deploy first-generation TDI - potentially offering simplifications in the processing (Section~\ref{Model}). 
  
  \item A lock acquisition scheme is proposed that address the transient behaviour when engaging the arm locking controller in the presence of Doppler shifts. This was a limiting factor for the applicability of previous arm locking schemes. In pre-stabilisation integration with arm locking~\cite{FCST}, the cavity control system uses a frequency-tunable or sideband locking PDH scheme, supported by the arm locking controller. This allows the Doppler pulling in the controller (up to maximum of 10 MHz) to persist due to the tunable frequency. In this paper, we utilise a standard PDH locking system with a fixed-length optical resonator and have stringent restrictions on the Doppler pulling. In Section~\ref{Doppler_data}, the laser frequency pulling is shown to be restricted to $\pm 20$ kHz, significantly smaller than the line-width of the cavity (100 kHz), maintaining the lock on the cavity. This level of frequency pulling may introduce a small, transient degradation in cavity noise performance, as offsetting the cavity error signal from resonance can couple intensity noise into the readout scheme - but this noise will be suppressed by the arm locking control system. To achieve a suitable level of frequency pulling at start up, we need to estimate the Doppler shifts with high accuracy and populate them in the arm locking control system.
  
  \item  
  Two estimation methods for high accuracy estimates of Doppler shifts using on-board measurements have been proposed in Section IV; 1) averaging the inter-spacecraft link phase readout  and 2)using  the PRN ranging measurements. For a thermal-noise-limited cavity, the Doppler parameter estimates can be obtained after an integration time of at least 1200~s. If the cavity is at requirement-level performance, the integration time needed is 50,000~s. PRN measurements can be used for estimation, requiring at least 6300~s of data. These integration times are estimates based on the residual noise spectra and will require further analysis.
  
  \item The scheme proposed here is compatible with the LISA baseline design, requiring firmware updates for the laser control system with minimal or no changes to the hardware. Previous work involving pre-stabilisation integration with arm locking have been described in~\cite{FCST,exp_mdual}, and can provide more suppression compared to the combination in this paper. But compared to these previous approaches, which require hardware changes to the optical cavity or locking system, the proposed control scheme may require minor changes for the summation of the feedback electronics of both the sensors depending on the final implementation of the system. We expect the digital resources required would be larger than a simple cavity locker or phase locker, but by a small factor (less than a factor of 2 larger in our initial estimate).
  
\end{enumerate}

Lastly, the Doppler estimation described here are indicative only, and thus we acknowledge possible estimation methods involving complex control systems. For example, a scheme utilizing the PDH sensors could be envisioned to iteratively suppress visible pulling during lock acquisition phase, at the expense of possible 'in-band' perturbations. However the time-constants of such a scheme needs to be carefully analysed for stability and may require a thorough understanding of the LISA orbits. With further research, such schemes could potentially serve as alternatives for lock acquisition for arm locking.

\section{\label{Conclusions}Conclusions}
This paper showed the control system dynamics and the Simulink model of a new method that combines the locking of both LISA arms and a Fabry-Pérot cavity to enhance the suppression of the laser frequency noise. From the noise budget, the noise coupling into the interferometer response is predominantly contributed by the requirement-level cavity noise, below which clock and spacecraft motion noise will be dominant.  The suppression of requirement-level cavity noise upto 3 orders, demonstrated herein is sufficient  for first-generation TDI to meet the requirements of LISA (Figure~\ref{fig:noise_budget}), thereby reducing the complexity of post-processing computations.\newline
This paper also proposes solutions to improve the Doppler shift estimation that is required for lock acquisition, by estimating the Doppler trends using inter-spacecraft link and/or inter-spacecraft ranging measurements. The sensor used in this paper comprises of simple architecture of common arm sensor, and fixed cavity. The future scope could be to explore more complex schemes of these two sensors~\cite{tunablespacer,thorpe,dualarm,mdual}, that should help in better suppression while maintaining low Doppler pulling. Experimental verification of this combination would be useful in solidifying the theoretical work.
\section*{Acknowledgement}
{The authors acknowledge that this research was conducted with support from the Australian Research Council Centre of Excellence for Gravitational Wave Discovery (OzGrav), through project number CE170100004. Kirk McKenzie’s contribution was partially supported by a contract from Jet Propulsion Laboratory, California Institute of Technology.

}

\appendix
\section{Noise Sources}
The noise sources, considered in this paper, are summarised as follows:
\begin{enumerate}
    \item{Laser frequency noise: The equation used for simulating the laser frequency noise in a free running NPRO laser can be given as~\cite{mdual} 
        \begin{equation}\label{N1}
            \nu_{L_i} (f)=\frac{30000\hspace{0.08cm} \textrm{Hz}}{f} \hspace{0.1cm} \textrm{Hz}/\sqrt{\textrm{Hz
            }}.
        \end{equation}
    Here i can be 1, 2 or 3, referring to the laser frequency noise in the laser sources in the individual spacecrafts. Although there are noises sourced from Spacecrafts 2 and 3, they get suppressed as a result of the high gain approximation of the transponders in the spacecrafts. As a result, in this paper, the notation $\nu_L$ will refer to $\nu_{L_1}$, as Spacecraft 1 is considered the primary spacecraft.
    From the model in Figure~\ref{Model_n}, the output (at point C) due to the laser frequency noise can be written as
     \begin{equation}
    \begin{split}
       \nu_{C;L}(s)=&\nu_L(s) - \nu_{C;L}(s)G_1 (s)P_{+}(s)\\ & -\nu_{C;L}(s) G_2 (s)P_{\textrm{pdh}}(s),\\
\end{split}
       \end{equation}
     \begin{equation}\label{T1}
     \begin{split}
     LN(s) &= \frac{\nu_{C;L}(s)}{\nu_{L}(s)}\\ &=  \frac{1}{1+G_1 (s) P_{+} (s)+G_2(s) P_{\textrm{pdh}}(s) }
         \end{split}.
    \end{equation}
Here G$_1$(s) and G$_2$(s) are controllers of the arm sensor and the cavity sensor, respectively, described in Section~\ref{Controller_design}. It can be seen that both the sensors contribute to the suppression of the laser frequency noise.}
\item{ Shot noise: The shot noise occurs due to quantum fluctuations of the laser and is observed at the event when the laser strikes the photo detector. It can be modelled as shown below~\cite{mdual}:
	\begin{equation}
	   { \theta_{shot;ij}(s) = \sqrt{\frac{\hbar c}{2\pi}\frac{1}{\lambda P}}} \hspace{0.1cm} \textrm{cycles}/\sqrt{\textrm{Hz}}.
	\end{equation}
The equation refers to the shot noise arising due to laser from Spacecraft j being detected at Spacecraft i. As per the equation, if the laser has more optical power, P, the amount of shot noise would be lower and vice versa. Typical value for shot noise for the application of LISA can be estimated using the parameters and is found to be 6.9 $\mu$cycles/$\sqrt{\textrm{Hz}}$~\cite{LISAL3}.
\begin{equation}
            \theta_{shot;ij}(s)=6.9\hspace{0.1cm}\textrm{x}\hspace{0.1cm}10^{-6}\hspace{0.1cm} \textrm{cycles}/\sqrt{\textrm{Hz}},
\end{equation}
\begin{equation}\label{N2}
        \nu_{shot;ij}(s)=(6.9\hspace{0.1cm} \textrm{x}\hspace{0.1cm} 10^{-6}).s \hspace{0.1cm} \textrm{Hz}/\sqrt{\textrm{Hz}}.
\end{equation}
The contribution from each shot noise can be shown as
      \begin{equation}
        \begin{split}
            \nu_{C;shot}(s)=&-\nu_{shot;ij}(s)G_1(s) - \nu_{C;shot}G_1(s) P_{+}(s) -\\ &\nu_{C;shot}(s)G_2(s) P_{\textrm{pdh}}(s)
        \end{split},
        \end{equation}
        \begin{equation} \label{T2}
        \begin{split}
            SN(s)=&\frac{\nu_{C;shot}(s)}{\nu_{shot;ij}(s)}\\
            =&\frac{-G_1 (s) }{1+G_1 (s) P_{+} (s)+G_2(s) P_{\textrm{pdh}}(s)}
        \end{split}.
        \end{equation}
        }
Shot noise can be seen mainly in the frequency bands where arm is dominant but gets suppressed otherwise. However, the amount of shot noise that couples into the system is very low compared to other noise sources and can be seen as the limits of noise suppression.

\item{ Cavity Noise: For LISA, during pre-stabilisation period, the laser frequency noise contribution can be reduced to a residual amount if the cavity is made using low-loss mirrors using ULE or Zerodur~\cite{PPA,cavity_noise}. This residual noise serves as a limitation of the stability provided by the cavity. 
 \begin{equation}\label{N3}
    \nu_{cavity} (s)=30 \left(\sqrt{1+\left(\frac{2\hspace{0.08cm}\textrm{mHz}}{f}\right)^4}\right)\hspace{0.1cm} \textrm{Hz}/\sqrt{\textrm{Hz}}.
\end{equation}
        
The contribution due to the cavity noise can be shown as 
 \begin{equation}
    \begin{split}
        \nu_{C;cavity}(s)=&-\nu_{cavity}(s)G_2(s)P_{\textrm{pdh}}(s) \\& -\nu_{C;cavity}\left(G_1(s) P_{+}(s) + G_2(s) P_{\textrm{pdh}}(s)\right)
    \end{split},
\end{equation}
\begin{equation}\label{T3}
    \begin{split}
        TN(s)=&\frac{\nu_{C;cavity}(s)}{\nu_{cavity}(s)}\\=&\frac{-G_2 (s) P_{\textrm{pdh}} (s)}{1+G_1 (s) P_{+} (s)+G_2(s) P_{\textrm{pdh}}(s)}
    \end{split}.
\end{equation}
The transfer function for the cavity noise is given by TN(s), one can see the cavity noise is suppressed by the arm locking control system.
}
\item{Clock Noise : It is caused by the clock signal on board the spacecraft, that is utilised for phase meter measurements~\cite{mdual}, and is dependent on the beat note frequency. The following equations describe the clock noise in LISA.
 \begin{equation}
          C_{i}(f)=\frac{\Tilde{y_{i}}(f)}{2\pi f} \hspace{0.1cm} \textrm{cycles}/\sqrt{\textrm{Hz}}.
 \end{equation}
Here the value of $\Tilde{y_{i}}$ is 2.4 x ${10^{-12}}/{\sqrt{f}} $  $1/\sqrt{\textrm{Hz}}$ and represents the fractional fluctuations of the clock that is used. The value of i can be 1, 2 or 3, corresponding to the clock in each individual phasemeter from Spacecraft 1, 2 or 3.
\begin{equation}                           
    \phi_{clock;ij}(f)=\Delta_{ij}C_{i}(f)\hspace{0.1cm} \textrm{cycles}/\sqrt{\textrm{Hz}} ,
\end{equation}
\begin{equation}\label{N4}  
    \nu_{clock;ij}(f)=\Delta_{ij}.C_{i}(f).s \hspace{0.1cm} \textrm{Hz}/\sqrt{\textrm{Hz}}.
\end{equation}
The beat note frequency, $\Delta_{ij}$ between two Spacecrafts i and j, is given a value of 30 MHz, assuming the worst-case scenario of Doppler pulling between the spacecrafts to be 5 MHz ($\pm$5 MHz) and a maximum heterodyne measurement offset of 25 MHz between two spacecrafts~\cite{LISAL3}. The corresponding noise response due to clock noise can be computed as
    
\begin{equation}
    \begin{split}
         \nu_{C;clock}(s)=-&\nu_{clock;ij}(s)G_2(s) - \nu_{C;clock}G_1(s) P_{+}(s) \\- &\nu_{C;clock}(s)G_2(s) P_{\textrm{pdh}}(s),
    \end{split}
\end{equation}
\begin{equation} \label{T4}
    \begin{split}
            CN(s)=&\frac{\nu_{C;clock}(s)}{\nu_{clock;ij}(s)}\\=&\frac{-G_1 (s)}{1+G_1 (s) P_{+} (s)+G_2(s) P_{\textrm{pdh}}(s)}
    \end{split}.
\end{equation}
    }
    The clock noises would be correlated if the same clock source is used for phase meter measurement and uncorrelated if there are separate clock sources.
    
\item{Spacecraft Motion noise:
    This noise is generated when the spacecraft follows the proof masses to retain drag free operation and cause inter-spacecraft jitters and is given by:~\cite{exp_mdual}
\begin{equation}
        \Delta\Tilde{X_{ij}}(s) = 1.5\sqrt{1+\left(\frac{8\hspace{0.08cm}\textrm{mHz}}{f}\right)^4} \hspace{0.1cm} \textrm{nm}/\sqrt{\textrm{Hz}},
\end{equation}

\begin{equation}
        \phi_{SC;ij} (s)=\frac{\Delta\Tilde{X_{ij}}(s)}{\lambda}  \hspace{0.1cm} \textrm{cycles}/\sqrt{\textrm{Hz}},
\end{equation}
    
\begin{equation}\label{N5}
        \nu_{SC;ij} (s)=\phi_{SC;ij} (s). s \hspace{0.1cm} \textrm{Hz}/\sqrt{\textrm{Hz}},
\end{equation}
    where $\lambda$ is the laser wavelength and is 1064nm for LISA. The corresponding noise response due to spacecraft motion can be computed as
    
\begin{equation} \label{T5}
    \begin{split}
            SCN(s)=&\frac{\nu_{C;SC}(s)}{\nu_{SC;ij}(s)}\\=&\frac{-G_1 (s)}{1+G_1 (s) P_{+} (s)+G_2(s) P_{\textrm{pdh}}(s)}
    \end{split}.
\end{equation}
    
The spacecraft motion noise, $\nu_{SC;ij}$ refers to the jitter between Spacecraft i and j, to match the proof-mass in spacecraft i. Both the clock noise and spacecraft motion have the same transfer function and hence, reaches the limit in the LISA science band while getting suppressed when the cavity is dominant.
    
The total spacecraft motion noise propagated through the system can be shown as below:
\begin{equation}
\begin{split}
        \nu_{SC}(s)=&-\nu_{SC;12}(s)[1+e^{-2s\tau_{12}}]-2[\nu_{SC;21}(s) e^{-s\tau_{12}}]\\ &- \nu_{SC;13}(s) [1+e^{-2s\tau_{13}}]-2[\nu_{SC;31}(s)e^{-s\tau_{13}}].
\end{split}
\end{equation} }
\item {Digitization noise: Digital hardware used for phasemeters and controllers couple noise into the system, due to analog-digital conversions and precision of integer arithmetic. But the contribution is very small of the order of 10$^{-10}$ \cite{exp_mdual}, and is neglected for noise budget in this paper.}
\end{enumerate}

\section{Sinusoidal spacecraft separation orbital model}
With respect to the orbital dynamics of LISA that is described by the model in Equation~\ref{Doppler_eqn}, the frequency pulling at lock acquisition can be reduced further if the Doppler shifts are modelled as a combination of sinusoids of half-year and a year periods, rather than the polynomial model of Doppler shifts in Section~\ref{Doppler_data}. The sinusoidal model is:
\begin{equation}\label{Dopplerlock_sine}
    \nu_{D;est}(t)= \nu_{0;+} + \gamma_{0;+}t +  \iint\alpha(t)dt'dt,
\end{equation}
where 
\begin{equation}\label{T6b}
\alpha(t)= \hat{\alpha_1} \sin (\hat{\omega_1} t + \hat{\phi_1}) + \hat{\alpha_2} \sin(\hat{\omega_2} t +\hat{\phi_2}).
\end{equation}

Here $\nu_{0;+}$ and $\gamma_{0;+}$ are the estimates of the Doppler shift and the first derivative of Doppler shift (Doppler rate) at the instant when the controller is just turned on. $\hat{\alpha_1}$ and $\hat{\alpha_2}$ are amplitude estimates of the two sinusoids of frequency estimates $\hat{\omega_1}$ and $\hat{\omega_2}$, along with phase shift estimates $\hat{\phi_1}$ and $\hat{\phi_2}$, that provides the Doppler acceleration of the system.  

\begin{table*}
\centering
\caption{Parameter requirements for orbital knowledge in order to meet lock acquisition conditions with the controller shown in Fig~\ref{fig:open_loop}. Each parameter's error limit is checked in combination with the errors of other parameters in Monte Carlo simulations. The achievable levels are cross-checked with estimation using Fabry-Pérot (FP) cavity estimation in~\cite{mdual} and PRN ranging, or Thermal Noise Limited (TNL) cavity estimation as described in Section~\ref{Timescales}. The other parameters require more information using orbital analysis before or during the commissioning of LISA.}    
\label{Dopplerreq_sine}      
\begin{tabular}{|c|c|c|c|c|}     
\hline       
  Parameter  &  Actual/Max  & Max error & Fractional & Estimation\\  & Value  & tolerance ($\pm$) &change  &methods \\ 
\hline  
    $\nu_0$ & 12 MHz & 10 Hz  & 8.33 x $10^{-7}$  & FP cavity estimation/ TNL estimation\\
    $\gamma_0$ &   4 Hz/s & 60 $\mu$Hz/s  & 2.5 x $10^{-4}$ & PRN ranging/ TNL estimation\\
    $\alpha_1$ & -1 $\mu$Hz/s$^2$ & 20 nHz/s$^2$  & 2 x $10^{-2}$& PRN ranging/ TNL estimation\\
    $\alpha_2$ & 0.25 $\mu$Hz/s$^2$ & 20 nHz/s$^2$  & 8 x $10^{-2}$ & PRN ranging/TNL estimation\\
    $f_1$ &  63.4 nHz & 0.1 nHz & 1.57 x $10^{-3}$ &  Orbital dynamics\\
    $f_2$ &  31.7 nHz & 0.1 nHz & 3.15 x $10^{-3}$ &   Orbital dynamics\\
    $\phi_1$ & 2$\pi$ rad & 10 $\mu${rad}   & 1.6 x $10^{-6}$ &  Orbital dynamics \\
    $\phi_2$ & 2$\pi$ rad & 10 $\mu${rad}  & 1.6 x $10^{-6}$  &Orbital dynamics\\
\hline                  
\end{tabular}
\end{table*}
The  maximum allowed values for the error in the  Doppler shift parameters  are given in Table~\ref{Dopplerreq_sine} for the  sinusoidal orbital model. The  values were  obtained by Monte Carlo simulations that select outputs that maintain less than $\pm$20 kHz deviation of laser frequency at lock acquisition. The results of the simulations are shown in Figure~\ref{pulling_transient_sine}. Figure~\ref{Doppler_orbits_sine} shows the spread of the sinusoidal model, at the worst-case orbit phase, based on the parameters in Table~\ref{Dopplerreq_sine}.  
The sinusoidal model is more accurate than the polynomial model (but more complex). If parameters estimates for the orbital model are better than
\begin{figure}[H]
\hspace*{-0.25cm}
\includegraphics[width=6.5cm, height=9.5cm, angle=-90 ]{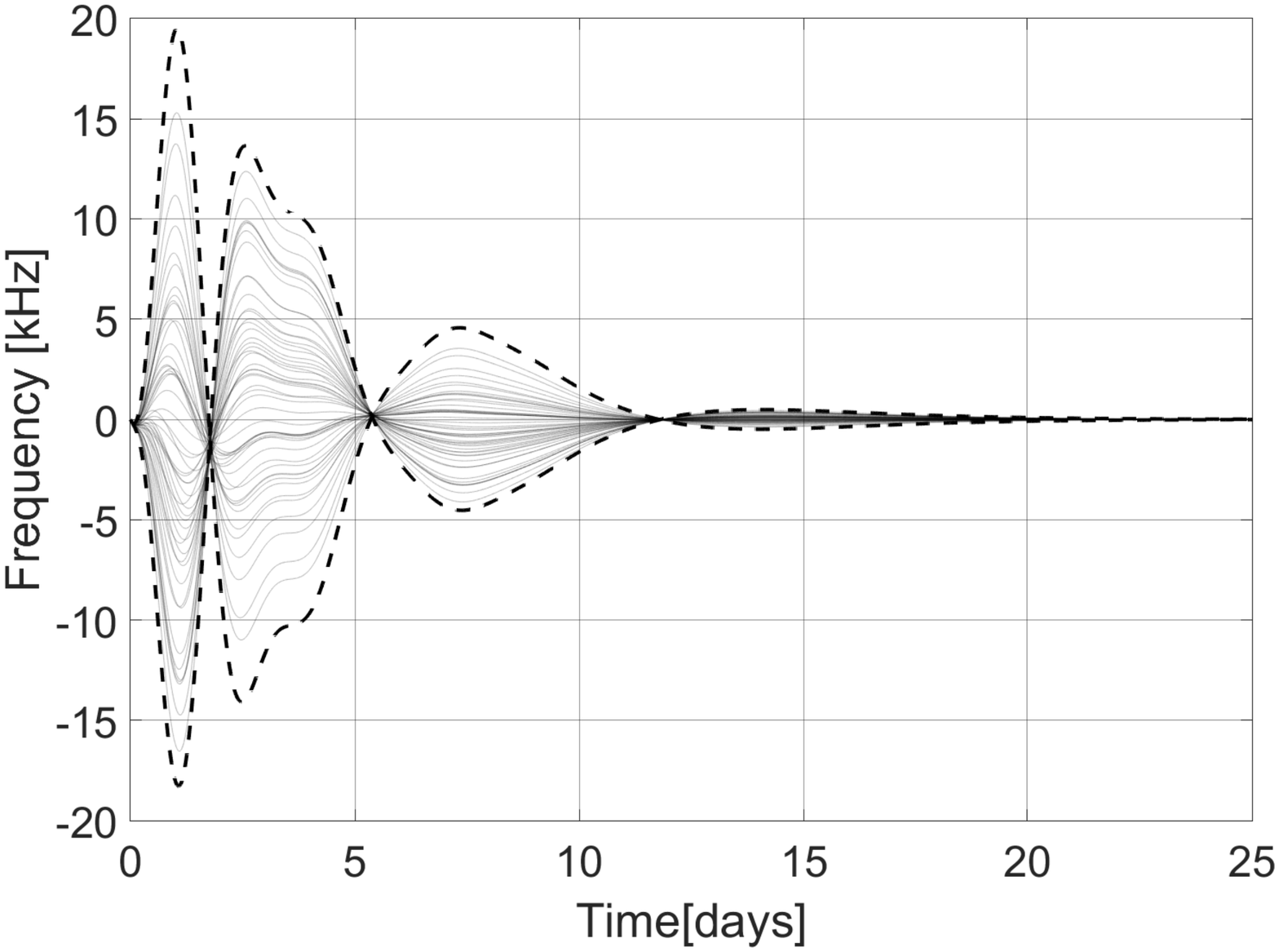}
\caption{Frequency deviation from cavity line center due to Doppler-error-pulling at lock acquisition. The estimate is calculated using the sinusoid functions in Equation~\ref{Dopplerlock_sine} for estimation. The black traces shows the different Monte Carlo simulations with the dashed traces showing the pulling with the  maximum error limits, given in the Table~\ref{Dopplerreq_sine}. The maximum pulling is 20 kHz and the HWHM linewidth of the cavity is 100 kHz, maintaining  the cavity lock.}
\label{pulling_transient_sine}
\end{figure}
those in Table~\ref{Dopplerreq_sine}, using the sinusoidal orbital model can lead to a smaller frequency deviation at lock acquisition than is possible to achieve with the polynomial model, which is limited by the intrinsic model error. Doppler shift parameter estimation shown in Section~\ref{Timescales} will be able to converge to the required precision of Table IV.

Alternatively, FP cavity estimation with LISA requirement-level performance, described in~\cite{mdual} can be used for the estimation for $\nu_0$ while the PRN ranging can be be used for estimating $\gamma_0$, $\hat{\alpha_1}$ and $\hat{\alpha_2}$, similar to the analysis in Section~\ref{Timescales}. We expect the parameters $\hat{f_1}$, $\hat{f_2}$, $\hat{\phi_1}$ and $\hat{\phi_2}$ to be derived from orbital models of LISA described analytically or measured during commission period. During the transient of the lock acquisition, the cavity is detuned up to 20 kHz. Though it is still in the linear regime, the optimal noise performance may be compromised to some extent, discussed earlier.
 
\begin{figure}[H]
\hspace*{-0.6cm}
\includegraphics[width=7cm, height=10cm, angle=-90 ]{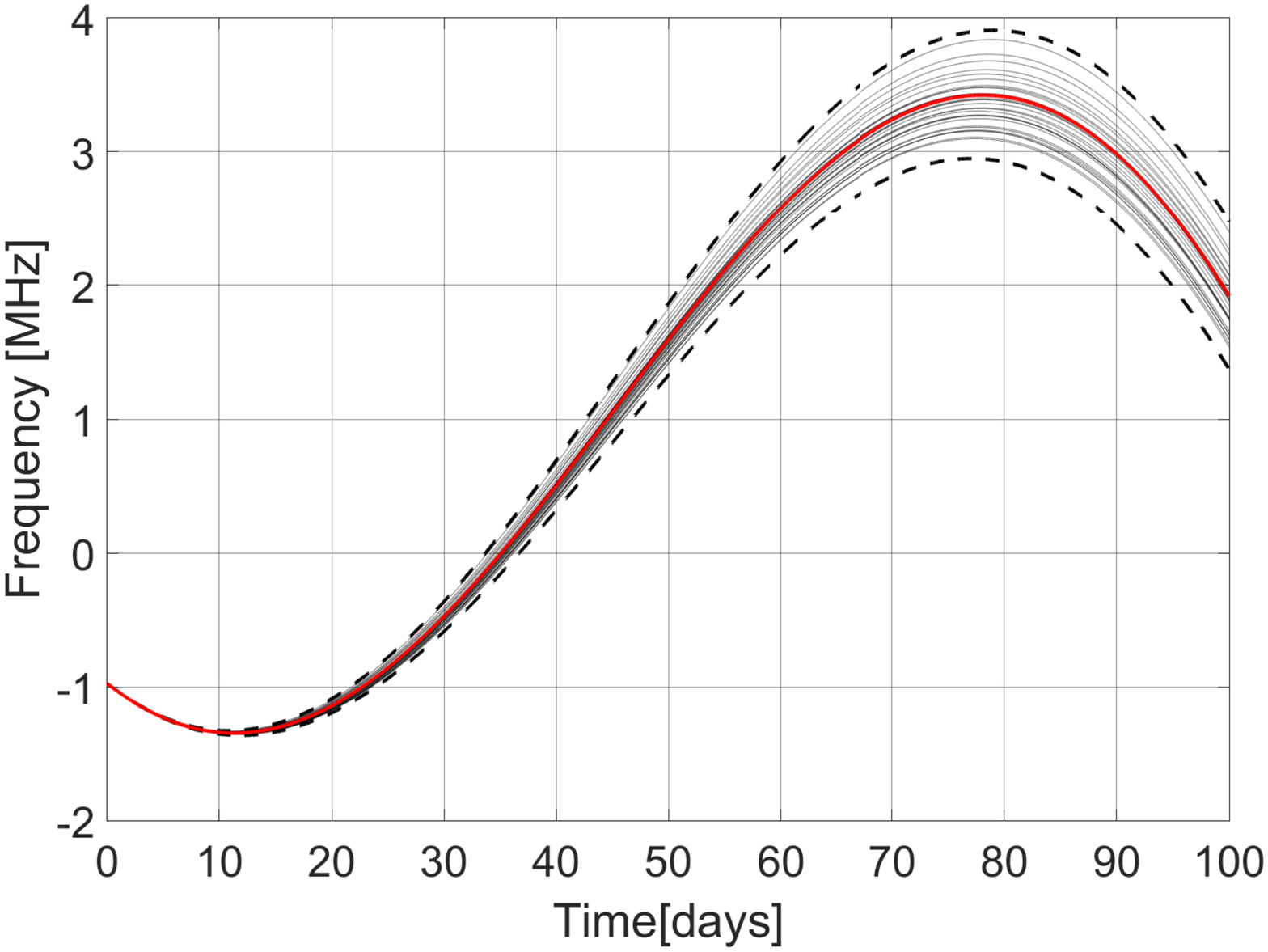}
\caption{Doppler shifts with the estimation models using the sinusoid functions in Equation~\ref{Dopplerlock_sine} for estimation. With the red dotted trace showing the actual Doppler shifts in Equation~\ref{Doppler_eqn}, the black traces shows the various Monte Carlo orbital models for different errors in the parameters with the dashed lines trace showing the worst case, given in the Table~\ref{Dopplerreq_sine}.}
\label{Doppler_orbits_sine}
\end{figure}

 \section{Implementation of Controllers}
 
 For ease of implementation,  stage I of  controller 1 was implemented as a sum of multiple low pass filters with appropriate gains. The poles and gains of the low pass structure are given in Table \ref{LP_table}
\begin{equation}
    G_{1;I}(s)=\frac{g_0}{s^2}\sum_{i=1}^{13}\frac{g_i}{s+p_i}.
\end{equation}
\begin{table}[H]

\centering
\caption{Parameters of the gains and poles for the Low-pass filter cascade to implement $1/{s^{0.3}}$ in Equation~\ref{G1} }           
\label{LP_table}      
\begin{tabular}{c c c}     
\hline\hline       
Index & Pole (p$_l$) & Gain(g$_l$) \\ 
number (l) & & \\
\hline           
\\
  0 &   & (2$\pi$ x 1.36 x $10^4$ )$^{2.3}$ x 12.31\\
  1 & 2$\pi$ x 5 x $10^{-6}$ & 2x $10^{-7}$ \\
  2 & 2$\pi$ x 1.6 x $10^{-4}$ & 2 x $10^{-4}$ \\
  3 & 2$\pi$ x 4 x $10^{-4}$ & 5 x $10^{-4}$ \\
  4 & 2$\pi$ x 5 x $10^{-3}$ & 5 x $10^{-3}$ \\
  5 & 2$\pi$ x 7.5 x $10^{-2}$& 3 x $10^{-2}$ \\
  6 & 2$\pi$ x 0.5 & 8 x $10^{-2}$ \\
  7 & 2$\pi$ x 4 & 0.5 \\
  8 & 2$\pi$ x 50 & 2.5 \\
  9 & 2$\pi$ x 400 & 12 \\
  10 & 2$\pi$ x 6000 & 110 \\
  11 & 2$\pi$ x 1.5 x $10^{5}$ & 1500 \\
  12 & 2$\pi$ x 2 x $10^{6}$ & 7000 \\
  13 & 2$\pi$ x $10^{-10}$ & $10^{-4}$ \\
  
\hline                  
\end{tabular}
\end{table}

 A similar approach was done to implement Controller 2 using the values in Table \ref{LP_table_2}
\begin{equation}
    G_{2}(s)=\frac{g_0}{s}\sum_{i=1}^{13}\frac{g_i}{s+p_i}.
\end{equation}
\begin{table}[H]
\caption{Parameters of the gains and poles for the Low-pass filter cascade to implement $1/{s^{0.5}}$ in Equation~\ref{G2} }
\label{LP_table_2}      
\begin{tabular}{c c c}     
\hline\hline       
Index & Pole (p$_l$) & Gain(g$_l$) \\ 
number (l) & & \\
\hline           
\\
  0 &   & 34 x (2$\pi$ x 7.32  x $10^3$ )$^{0.5}$\\
  1 & 2$\pi$ x 3 x $10^{-5}$ & 3.5 x $10^{-4}$ \\
  2 & 2$\pi$ x 3 x $10^{-4}$ & 8.82 x $10^{-4}$ \\
  3 & 2$\pi$ x 3 x $10^{-3}$ & 3.12 x $10^{-3}$ \\
  4 & 2$\pi$ x 3 x $10^{-2}$ & 8.82 x $10^{-3}$ \\
  5 & 2$\pi$ x 0.3 & 3.12 x $10^{-2}$ \\
  6 & 2$\pi$ x 3 & 8.82 x $10^{-2}$ \\
  7 & 2$\pi$ x 30 & 0.312 \\
  8 & 2$\pi$ x 3 x $10^{2}$ & 0.882 \\
  9 & 2$\pi$ x 3 x $10^{3}$ & 3.12 \\
  10 & 2$\pi$ x 3 x $10^{4}$ & 8.82 \\
  11 & 2$\pi$ x 3 x $10^{5}$ & 31.2 \\
  12 & 2$\pi$ x 3 x $10^{6}$ & 88.2 \\
  13 & 2$\pi$ x $10^{-10}$ & $10^{-4}$ \\
  
\hline                  
\end{tabular}
\end{table}

\end{document}